\documentclass[aps,preprint]{revtex4}
\usepackage{amsmath}
\usepackage{bm}
\usepackage{graphicx}
\newcommand{\ve}{\mathbf{e}}

\newcommand{\vk}{\mathbf{k}}

\newcommand{\vp}{\mathbf{p}}
\newcommand{\vq}{\mathbf{q}}

\newcommand{\hk}{\mathbf{\hat k}}
\newcommand{\hp}{\mathbf{\hat p}}
\newcommand{\hq}{\mathbf{\hat q}}

\newcommand{\vy}{\mathbf{y}}

\newcommand{\hx}{\mathbf{\hat x}}
\newcommand{\hy}{\mathbf{\hat y}}

\newcommand{\vvp}{\vec p}
\newcommand{\var}{\vec r}
\newcommand{\vvk}{\vec k}

\newcommand{\deriv}{\mathop\partial\nolimits}

\begin{document}

\title{Multipole expansions and Fock symmetry of the Hydrogen atom}

\author{A.~V.~Meremianin}
\email{avm@pks.mpg.de}
\affiliation{Max-Planck-Institute for the Physics of Complex Systems,
  N\"othnitzer Str. 38, 01187, Dresden, Germany}
\affiliation{REC-010, Voronezh State University, 394006, Voronezh, Russia}

\author{J-M.~Rost}
\email{rost@pks.mpg.de}

\affiliation{Max-Planck-Institute for the Physics of Complex Systems,
  N\"othnitzer Str. 38, 01187, Dresden, Germany}
\date{\today}

\begin{abstract}
The main difficulty in utilizing the $O(4)$ symmetry of the Hydrogen atom in
practical calculations is the dependence of the Fock stereographic projection
on energy.
This is due to the fact that the wave functions of the states with different
energies are proportional to the hyperspherical harmonics (HSH) corresponding
to different points on the hypersphere.
Thus, the calculation of the matrix elements reduces to the problem of
re-expanding HSH in terms of HSH depending on different points on the
hypersphere.
We solve this problem by applying the technique of multipole expansions
for four-dimensional HSH.
As a result, we obtain the multipole expansions whose coefficients
are the matrix elements of the boost operator taken between
hydrogen wave functions (i.e. hydrogen form-factors).
The explicit expressions for those coefficients are derived.
It is shown that the hydrogen matrix elements can be presented as derivatives
of an elementary function.
Such an operator representation is convenient for the derivation of recurrency
relations connecting matrix elements between states corresponding to different
values of the quantum numbers $n$ and $l$.
\end{abstract}

%% \pacs{33.15.Mt,33.20.Sn}

\maketitle

%%%%%%%%%%%%%%%%%%%%%

\section{Introduction}
\label{sec:introduction}

About seventy ears ago, Fock in his paper \cite{fock35:_o4} demonstrated
that the wave functions of the hydrogen atom in momentum representation are
proportional to the four-dimensional hyperspherical harmonics (HSH).
From the point of view of the group theory, it means that the hydrogen wave
functions possess the $O(4)$ symmetry. 
The source of this remarkable property of the quantum two-body Coulomb problem
lies in the presence in of an additional conserving vector quantity -- the
Laplace-Runge-Lenz operator.

The applications of the Fock's method to various problems of quantum
mechanics are very extensive.
Lieber used it to calculate the Lamb shift in the Hydrogen atom
\cite{lieber68:_o4_lamb}.
Group-theoretical properties of $O(4)$ symmetry of the hydrogen atom have been
analyzed in \cite{bander_itzykson66I:_o4_hydrog}.
Schwinger \cite{schwinger64:_c_green_f} showed that the hydrogen Green
function can be presented in a simple form as the series of four-dimensional
hyperspherical harmonics.
Fock's method has been used for the calculation of the retardation effects in
two-photon bound-bound transitions \cite{maquet77:_green_fock}.
The four-dimensional HSH can be chosen as a Sturmian basis set in the
many-center Coulomb problem
\cite{shibuya_wulfman65:_mo_fock,avery04:_shibuya_wulfman_fock,%
avery00:_hspher_book,aquilanti98prl:_o4_recouplings,AquilantiCCC03,%
aquilanti01:_rev_hypsp_mom_space,AquilantiCC97,AquilantiCCG96,%
AveryA04,AveryS01,avery00:_hspher_book,Avery97,AveryH96,AveryHWA96}.
Recently, the Fock's projection method has been applied to the theory of
anisotropic excitons, see \cite{muljarov00jmp:_exciton_fock}.

As was pointed by Lieber \cite{lieber68:_o4_lamb}, the main difficulty in
applying Fock's symmetry to the specific problems is the dependence of the
arguments of HSH on the energy.
This prohibits the direct application of the Wigner-Eckart theorem to the
calculation of matrix elements between different states of the
two-body Coulomb system.
Often, this problem can be avoided by using the above mentioned set of
Sturmian functions.
Those functions (denoted sometimes as ``Sturmians'') have the same form as the
bound-state wave functions of the hydrogen atom whose arguments are chosen to
be equal. 
Thus, it is possible to apply the technique of the angular momentum theory to
the calculation of integrals involving Sturmian functions which belong to the
same set \cite{AquilantiC02,avery00:_hspher_book}.

If one needs to calculate integrals between Sturmians from different sets,
there appears the same problem as in the calculation of integrals involving
hydrogen wave functions with different energies.
Namely, the Wigner-Eckart theorem can be applied only for integrals containing
HSH whose arguments correspond to the same point on the hypersphere.
However, hydrogen wave functions with \textit{different} energies (or
Sturmians from different sets) correspond to \textit{different} points on the
hypersphere.
In this paper we show that this problem can be solved using the multipole
expansion theorems for four-dimensional HSH.
Mathematically, we have derived the connection between sets of
HSH corresponding to different stereographic projections.
That is, HSH depending on some point of the hypersphere can be presented as a
series of HSH depending on some other point which corresponds to the different
stereographic projection of the three-dimensional space onto the
four-dimensional hypersphere.
The coefficients in those expansions are, in fact, the form-factors of the
Hydrogen atom.

The form factor integrals are also known as ``generalized oscillator
strengths'' (GOS).
GOS play an important role in the theory of collisions of fast particles with
atoms and molecules. 
Starting from the pioneering work by Bethe
\cite{bethe30:_GOS_passage_rapid_particles}, there is a large number of papers
addressing the properties of GOS of the hydrogen atom (see review
\cite{inokuti71:_GOS_Bethe_inelastic}).
In the book \cite{mott_massey65} the compact expressions for GOS
corresponding to the transitions from $1s$ and $2s$ to $n,l$ states are given.
The hydrogenic form-factors for transitions between arbitrary bound states
were presented in \cite{jetzke86:_boost_hydrogen} in terms of combinations of
hypergeometric functions of two variables.
The integrals between the hydrogen wave functions and the operator of the
electromagnetic interaction have been calculated in
\cite{parzynski03:_h_mel_nondip} using the set of parabolic coordinates.
The expressions for the bound-free matrix elements for the hydrogen atom may
be found, e.g.~in
\cite{holt69:_hydrogen_mel_bound_free,datta85:_coulom_integrals}.
In all above works the calculations have been performed by means of
straightforward evaluation of the three-dimensional integrals in spherical or
parabolic coordinates.

Below we demonstrate that the hydrogen form-factor integrals can be explicitly
calculated \textit{without} evaluating any integrals.
Our approach is based solely on the addition theorems for four-dimensional HSH
which have been derived using the differential technique
\cite{avm05:_o4_jpa}.
As a result, we can write the hydrogen form-factor integrals as a combination
of multipole terms (they are equivalent to the radial integrals).
Each multipole term can be presented in a compact closed form containing
derivatives of a some elementary function.
This differential representation is very convenient for analyzing various
properties of the integrals such as their asymptotic behavior.
Moreover, using the differential representations we have derived a
number of recursion relations connecting radial integrals corresponding to
different values of their indices.

The paper is organized as follows.
In sec~\ref{sec:ster-proj} we consider the connection between matrix elements
in coordinate and momentum representations as well as the general properties
of the stereographic projection.
The expression for the matrix element of the boost operator in terms of HSH
is derived in sec.~\ref{sec:matr-elem-moment}.
We demonstrate that the calculation of the form-factor integral reduces to
the problem of the derivation of a certain multipole expansion for the product
of HSH with some scalar function.
Then, in sec.~\ref{sec:prop-ster-proj}, we use Fock's method of stereographic
projections in order to analyze the properties of the function whose multipole
expansion has to be derived.

In section~\ref{sec:mult-expans-transl}, we calculate such multipole
expansion.
First, in sec.~\ref{sec:mult-expans-terms}, the multipole expansion is derived
in the tensor form, i.e. as a tensor product of HSH.
Next, in sec.~\ref{sec:calc-mult-coeff}, we calculate the multipole
coefficients explicitly.
In Section \ref{sec:mult-expans-transl} the properties of the matrix elements
are considered regarding the expressions for the multipole coefficients.
In sec.~\ref{sec:dipole-limit} the dipole limit of the matrix elements is
analyzed in detail.
Some concluding remarks are given in Sec.~\ref{sec:conclusion}.
Appendix~\ref{sec:calculation-sum-over} contains the calculation of
summations arising in the course of computation of the multipole
coefficients.
In Appendix~\ref{sec:expl-P-params} we derive explicit expressions for the
multipole coefficients.
Recurrency relations for the multipole coefficients are considered in
Appendix~\ref{sec:recurrences}.

Below we use bold font letters like $\vp$, $\vk$ to denote vectors in
four-dimensional space.
For unit vectors in that space the hat is used, e.g. $\hx$, $\hy$.
Three-dimensional vectors are denoted as $\vvp$, $\vvk$, \textit{etc}.

%%%%%%%%%%%%%%%%%%%%%%%%%

\section{Stereographic projections and matrix elements}
\label{sec:ster-proj}

In this section we present the relation between hydrogen wave function in
coordinate and momentum representations.
This is done in sec.~\ref{sec:matr-elem-moment} where also the expression
for the matrix element of the boost operator in terms of HSH is derived.
Next, in sec.~\ref{sec:prop-ster-proj} we discuss the general properties of
the Fock's stereographic mappings corresponding to HSH entering the
form-factor integral.

%%%%%%%%%%%%%%%%%%%%%

\subsection{The matrix elements in the momentum space}
\label{sec:matr-elem-moment}

The hydrogen wave function in momentum representation $\psi_{nlm} (\vvp)$ is
defined as the Fourier transform of the wave function in coordinate
representation,
\begin{equation}
  \label{eq:c-1}
  \begin{split}
      \psi_{nlm} (\vvp) &= \frac{1}{(2\pi)^{3/2}} 
 \int e^{- i \,\vvp \cdot \var}
 \psi_{nlm} (\var)\, d \var, \\
  \psi_{nlm} (\var) &= \frac{1}{(2\pi)^{3/2}} \int e^{ i \,\vvp \cdot \var}
 \psi_{nlm} (\vvp) \, d \vvp,
  \end{split}
\end{equation}
where $n$ is the principal quantum number, $l,m$ are the angular momentum
quantum numbers, $l=0,1,\ldots n-1$ and  $m=-l,-l+1,\ldots l$.
The explicit expression for the wave function in momentum space has the
form \cite{fock35:_o4},
\begin{equation}
  \label{eq:h-1}
  \psi_{nlm} (\vvp) = \frac{4 \beta^{5/2}}{(\beta^2+(\vvp)^2)^2} 
Y_{n-1\,lm} (\hx), \quad \beta=Z/n.
\end{equation}
Here, $Y_{n-1\,lm} (\hy)$ is the four-dimensional HSH and 
$\hy$ is the unit $4$-vector whose components are defined by
\begin{equation}
  \label{eq:xi-1}
    \hy =
    \left(
      \frac{2\beta \vvp}{\beta^2 + (\vvp)^2},
      \frac{\beta^2 - (\vvp)^2}{\beta^2 + (\vvp)^2}
    \right).
\end{equation}
Functions $Y_{n l m}$ are normalized to the unity
\begin{equation}
  \label{eq:h-2}
  \int Y^*_{nlm} (\hy) Y_{n'l'm'} (\hy)  \, d \Omega_{\vy}
= \delta_{n,n'} \delta_{l,l'} \delta_{m,m'}.
\end{equation}
Here, it is supposed that the parameters $\beta$ entering the definition of
$\hy$ do not depend on indices $n,n'$.
The connection of the surface element of the four-dimensional hypersphere 
$d \Omega$ with the three-dimensional volume element $d \vvp$ has the form
\begin{equation}
  \label{eq:s-p}
  d \Omega_{\vy} = \left( \frac{2\beta}{\beta^2+(\vvp)^2} \right)^3 \,
 d \vvp.
\end{equation}
Note that, in the case of hydrogen, the vector $\hy$ depends on the rank $n$ of
HSH (see definition (\ref{eq:h-1}) of the parameter $\beta$).
The dependence of $\beta$ on the principal quantum number $n$ explains why the
orthogonality relations cannot be directly applied to the calculation of
matrix elements.

We remark that the method of Sturmian functions has found wide use in various
problems of the quantum chemistry and the theory of interaction of photons
with atoms and molecules.
In contrast to the set of the bound-state wave functions of the hydrogen atom,
the set of Sturmian functions is complete.
Sturmians in momentum representation are defined by eq.~(\ref{eq:h-1}) in
which $\alpha$ must be a fixed number independent of $n$.
Thus, the parameter $\alpha$ is the same for every function from the same
Sturmian set.

In many problems of the collision theory it is necessary to calculate the
form-factor integral,
\begin{equation}
  \label{eq:form-fact-1}
  F (\vvk) = 
\int \psi^*_{n_f} (\var)\, e^{i\, \vvk \cdot \var}\, \psi_{n_i} (\var)\, 
d \var,
%% \int \psi^*_{n_f} (\vvp) \psi_{n_i} (\vvp-\vvk) \, d \vvp,
\end{equation}
where, for the sake of shortness, we have omitted the angular momentum indices
which are $l'm'$ and $lm$ for the final and initial states $n_f$ and $n_i$,
respectively.
Using above eqs.~(\ref{eq:c-1}) the matrix element $F(\vvk)$ can be expressed
in terms of momentum space functions $\psi(\vvp)$ as
\begin{equation}
  \label{eq:mel-2}
F(\vvk) = \frac{1}{(2\pi)^{3}}
\int d \var \int d \vvp_1 \int d \vvp_2\,
 \psi^*_{n_f} (\vvp_1)\, e^{-i\, \vvp_1 \cdot \var}
e^{i\, \vvk \cdot \var}\,
\psi_{n_i} (\vvp_2)\, e^{i\, \vvp_2 \cdot \var}.
\end{equation}
Using the definition of the three-dimensional Dirac $\delta$-function, one
can evaluate integrals over $\vvp_2$ and $\var$ in closed form which yields
\begin{equation}
  \label{eq:mel-3}
F(\vvk) =
 \int \psi^*_{n_f} (\vvp) \, \psi_{n_i} (\vvp-\vvk)
 \, d \vvp.
\end{equation}
Using explicit expressions (\ref{eq:h-1}) and (\ref{eq:s-p}) for the wave
functions and introducing the indices $n=n_i-1$ and $n'=n_f-1$, the
form-factor becomes
\begin{equation}
  \label{eq:mel-4}
  F (\vvk) \equiv F_{n'l',nl} = 
\frac{2 \alpha^{5/2}}{\beta^{1/2}}
\int Y^*_{n'l'm'} (\hy)
\frac{\beta^2+(\vvp)^2}{(\alpha^2+(\vvp-\vvk)^2)^2}
 Y_{nlm} (\hx) \,  d \Omega_\vy,
\end{equation}
where the parameters $\alpha$ and $\beta$ have the form
\begin{equation}
  \label{eq:mel-5}
  \alpha = \frac{Z}{n_i} = \frac{Z}{n+1}, \quad
  \beta = \frac{Z}{n_f} = \frac{Z}{n'+1},
\end{equation}
and the components of the unit vector $\hx$ are defined by
\begin{equation}
  \label{eq:y-1}
      \hx =
    \left(
      \frac{2\alpha (\vvp-\vvk)}{\alpha^2 + (\vvp-\vvk)^2},
      \frac{\alpha^2 - (\vvp-\vvk)^2}{\alpha^2 + (\vvp-\vvk)^2}
    \right).
\end{equation}

%%%%%%%%%%%%%%%%%%%%%%%

\subsection{The properties of the stereographic projection}
\label{sec:prop-ster-proj}

\begin{figure}[htbp]
  \centering
  \includegraphics[width=5cm]{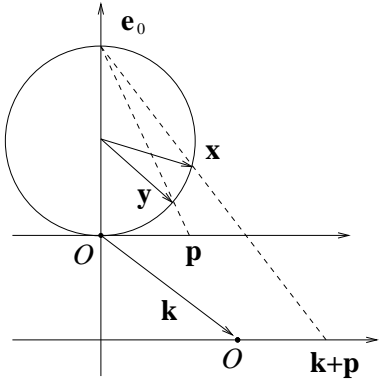}
  \caption{The cut through the four-dimensional hypersphere. 
  The three-dimensional $\vvp$-space corresponds to the horizontal axes.
  The vertical $z_0$-axis (the energy axis) is denoted as $\ve_0$, the
  point $O$ corresponds to the origin, $\vvp=0$. 
  The point $\hx$ on the hypersphere is the stereographic projection of 
  the point $\vp$ shifted by the vector $\vvk$.}
  \label{fig:stereo-1}
\end{figure}

The geometrical meaning of the unit vectors $\hy$ and $\hx$ in
eq.~(\ref{eq:mel-4}) and their connection to the three-dimensional vectors
$\vvp$ and $\vvk$ is demonstrated on Fig.~\ref{fig:stereo-1}.
It is seen that there is one-to-one correspondence between the
three-dimensional space of vectors $\vvp$ and the unit four-dimensional
hypersphere.
Namely, every point $\hx$ on the hypersphere corresponds to only one vector
$\vvp$ and vice versa.
This $\hx \leftrightarrow \vvp$ mapping is called stereographic
projection of the hyperplane (which is, in fact, the whole three-dimensional
space) on the hypersphere of unit radius.
%%%%%%%%%%
%%%%%%%%%%

As follows from eq.~(\ref{eq:mel-4}), in order to calculate the form-factor
integral one has to expand the product
\begin{equation}
 \label{eq:prod-1}
 \frac{\beta^2+(\vvp)^2}{(\alpha^2+(\vvp-\vvk)^2)^2}\, Y_{nlm} (\hx)
\end{equation}
in terms of HSH depending on the vector $\hy$.
When done so, the integral $F(\vvk)$ calculates trivially using the
orthogonality property of HSH.

As a preliminary, we note that vectors $\hy$ and $\hx$ can be re-written in
the compact form as
\begin{equation}
  \label{eq:xi-4}
   \hx = - \ve + 2 \alpha \frac{\vq}{q^2}, \quad  
   \hy = - \ve + 2 \beta \frac{\vp}{p^2},  \quad\vq=\vp-\vk
\end{equation}
where $\ve$ is the unit vector with components $\ve = (0,0,0,1)$ and 
$4$-vectors $\vp$ and $\vk$ are defined by
\begin{equation}
  \label{eq:xi-3}
  \vp = (\vvp,\beta), \quad \vk = (\vvk, \beta-\alpha).
\end{equation}
With new notations, the expression (\ref{eq:mel-4}) for the form-factor integral
assumes the form
\begin{equation}
  \label{eq:f-fact-1}
F_{n'l',nl} = \frac{2 \alpha^{5/2}}{\beta^{1/2}} 
\int Y^*_{n'l'm'} (\hy) \frac{p^2}{q^4} Y_{nlm} (\hx) \,d \Omega_\vy.
\end{equation}
Below, for simpler presentation, we will use the renormalized HSH 
$C_n (\hy)$ connected to the normalized HSH by
$C_{nlm} (\hy) = (-1)^{n-l} \sqrt{2 \pi^2/(n+1)}\, Y_{nlm}(\hy)$.
We will also omit the projection indices $l,m$ in the notation of HSH
$C_n$, so far this will not lead to misunderstandings.

The possibility of the efficient use of the multipole expansion technique for
calculation of integrals stems from the surprising fact that both the
products $p^{-2} C_n (\hy)$ and $|\vp-\vk|^{-2} C_n (\hx)$ satisfy the
four-dimensional Laplace equation with respect to the vector $\vp$.
Namely,
\begin{equation}
  \label{eq:lpl-1}
  \Delta_\vp \frac{1}{p^2} C_n (\hy) = 
\left( \Delta_{\vvp}
%%%% \frac{\partial^2}{\partial \vvp^2} 
+\frac{\partial^2}{\partial p_0^2} 
\right) \frac{1}{p^2} C_n (\hy) = 0, 
\quad
  \Delta_\vp \frac{1}{|\vp-\vk|^2} C_n (\hx) =0,
\end{equation}
where the vector $\vp$ has the components $\vp=(\vvp,p_0)$ and the unit
vectors $\hy$ and $\hx$ are defined by eqs.~(\ref{eq:xi-4}) in which $\beta$
and $\alpha$ are constants.
Note that equations (\ref{eq:lpl-1}) are valid for arbitrary values of $\vk$.

The equations (\ref{eq:lpl-1}) can be proved using the multipole expansions
formula derived in \cite{avm05:_o4_jpa},
\begin{equation}
  \label{eq:m-1}
  \frac{1}{p^2} C_{n} (\hy)
= \frac{2^{n/2}}{p^2}
\left\{
  -\ve + 2 \beta \frac{\vp}{p^2}
\right\}_n 
= \frac{1}{p^2}
\sum_{n_1=0}^n \binom{n}{n_1} (-1)^{n-n_1}
\left( \frac{2 \beta}{p} \right)^{n_1}
 \{ C_{n-n_1}(\ve) \otimes C_{n_1} (\hp) \}_n.
\end{equation}
Here the curly brackets mean the four-dimensional tensor products of the rank
$n$ which is the generalization of the conventional three-dimensional tensor
product \cite{Man-96,meremianin03:_phys_rep}.
 The product $p^{-n_1-2} C_{n_1} (\hp)$ which occurs on rhs of
eq.~(\ref{eq:m-1}) is the irregular solution of the Laplace equation. 
Therefore, the action of the Laplace operator $\Delta_\vp$ on
eq.~(\ref{eq:m-1}) gives zero,
\begin{equation}
 \label{eq:m-2}
  \Delta_\vp \frac{1}{p^{n_1+2}} C_{n_1} (\hp) =0.
\end{equation}
This completes the proof of the first equation in (\ref{eq:lpl-1}).
The second equation can be proved analogously noting that
\begin{displaymath}
  \Delta_\vp F(\vp-\vk) = \Delta_\vq F(\vq),
\end{displaymath}
where $F$ is an arbitrary function and $\vq=\vp-\vk$.

%% The product (\ref{eq:prod-1}) can be re-written in terms of vectors $\vq$ and
%% $\hx$ as,
%% \begin{equation}
%%   \label{eq:xi-5}
%%  \frac{1}{(\alpha^2+(\vvp-\vvk)^2)^2}\, C_{n} (\hx)
%% = \frac{1}{q^4}\, C_{n} (\hx).
%% \end{equation}
As is seen from eq.~(\ref{eq:lpl-1}), it is convenient to work with the product
$q^{-2} C_n (\hx)$ rather than with $q^{-4} C_n (\hx)$.
In order to establish the connection between these two combinations we employ
the auxiliary identity
\begin{displaymath}
\frac{1}{q^4}  = \frac{1+(\hx \cdot \ve)}{2\alpha^2} \frac{1}{q^2} 
\end{displaymath}
which can be proved by taking the square of the second equation in
(\ref{eq:xi-4}) and noting that $\ve^2=\hx^2=1$.
Applying the Clebsch-Gordan expansion technique one can prove the relation
\begin{multline}
  \label{eq:xi-6}
  (\hx \cdot \ve) C_{nlm} (\hx) = \frac{1}{\sqrt2}\, C_{100} (\hx) 
C_{nlm} (\hx) \\
= \frac{1}{\sqrt2}\, C^{(n-1) lm}_{100;\, nlm} C_{(n-1) lm} (\hx)
+ \frac{1}{\sqrt2}\, C^{(n+1) lm}_{100;\, nlm} C_{(n+1) lm} (\hx),
\end{multline}
where $C^{(n\pm 1) lm}_{100;\,nlm}$ is the four-dimensional Clebsch-Gordan
coefficient.
At such particular values of parameters, the explicit expressions for those
coefficients are given in ref.~\cite{avm05:_o4_jpa}.
As a result, the product $q^{-4} C_n (\hx)$ can be written as
\begin{multline}
  \label{eq:xi-7}
 \frac{1}{q^4}\, C_{nlm} (\hx)
= \frac{1}{ 2 \alpha^2\, q^2} 
 \Biggl(
 C_{nlm} (\hx) + \frac{\sqrt{(n-l+1)\,(n+l+2)}}{2 (n+1)} C_{(n+1) lm}(\hx) \\
%%%%%%%%%%%
     + \frac{\sqrt{(n-l)\,(n+l+1)}}{2 (n+1)} C_{(n-1) lm}(\hx)
 \Biggr).
\end{multline}
Thus, the problem reduces to the calculation of the multipole decomposition of
the function $q^{-2} C_n (\hx)$.
Once such the decomposition is known, the expression for the function
$q^{-4} C_n (\hx)$ can be derived using eq.~(\ref{eq:xi-7}).

%%%%%%%%%%%%%%%%%%%%%%%%%%

\section{The multipole expansion of translated harmonics}
\label{sec:mult-expans-transl}

In this section we derive the expansion of the function
$q^{-2} C_n (\hx)$ in terms of HSH depending on the unit vector $\hy$.
In sec.~\ref{sec:mult-expans-terms} the tensor form of the multipole expansion
is derived by consequent application of the multipole series given in
\cite{avm05:_o4_jpa}.
In sec.~\ref{sec:calc-mult-coeff} the coefficients of that expansion are
calculated in an explicit form.

%%%%%%%%%%%%%%%%%%%%%%

\subsection{The multipole expansion in terms of tensor products}
\label{sec:mult-expans-terms}

As was noted above, it is sufficient to calculate the expansion of the
product $q^{-2} C_{n} (\hx)$.
%% Then, eq.~(\ref{eq:xi-7}) can be used to obtain the expression for the
%% coefficients of the expansion of $q^{-4} C_n (\hx)$.
The calculations are performed in three steps.
First, we expand the product $q^{-4} C_n (\hx)$ in terms of HSH depending on
the hyper-angles of the vector $\vq$.
Second, the resulting HSH is expanded in terms of HSH whose arguments are the
hyper-angles of the $4D$-momentum vector $\vp$.
The final, third step, is to expand HSH depending on $\vp$ in terms of the
HSH depending on the unit vector $\hy$.
Clearly, at each step the series of irreducible tensor products of HSH
will occur.
However, as we will see below, only one of the three series is infinite.
Further, the Clebsch-Gordan coefficients inherent to the irreducible tensor
products arising during the computations have simple structure and can be
written in a closed form.

We begin the calculations by using the expansion similar to
eq.~(\ref{eq:m-1}),
\begin{equation}
  \label{eq:m-3}
\frac{1}{q^2} C_{n} (\hx) 
= \frac{1}{q^2}
\sum_{n_1=0}^n \binom{n}{n_1} (-1)^{n-n_1}
\left( \frac{2 \alpha}{q} \right)^{n_1}
 \{ C_{n-n_1}(\ve) \otimes C_{n_1} (\hq) \}_n.
\end{equation}
Now, we have to re-expand the function $q^{-n_1-2} C_{n_1}(\hq)$ in terms
of HSH depending on the vector $\vp$.
The use of the multipole expansion formula (eq.~(65) of
ref.~\cite{avm05:_o4_jpa}) yields,
\begin{equation}
  \label{eq:m-4}
  \frac{1}{q^{n_1+2}} C_{n_1} (\hq)
= \sum_{n_2=0}^\infty \frac{k^{n_2}}{p^{n_1+n_2+2}}
\binom{n_1 + n_2+1}{n_2} 
 \{ C_{n_2}(\hk) \otimes C_{n_1+n_2} (\hp) \}_{n_1}
\end{equation}
Third, we have to go back from vectors $\vp$ to vectors $\hy$.
In order to do that we note that the terms containing vector $\vp$ can be
re-written as
\begin{equation}
  \label{eq:m-5}
  \frac{1}{p^{n_1+n_2}} C_{n_1+n_2} (\hp)
= \frac{2^{(n_1+n_2)/2}}{p^{2n_1+2n_2}} \{ \vp \}_{n_1+n_2}
= 2^{(n_1+n_2)/2}
  \left\{
    \frac{\vp}{p^2}
  \right\}_{n_1+n_2}.
\end{equation}
The connection (\ref{eq:xi-4}) between vectors $\vp$ and $\hy$ leads to the
chain of equations
\begin{multline}
  \label{eq:m-6}
  2^{(n_1+n_2)/2}
  \left\{
    \frac{\vp}{p^2}
  \right\}_{n_1+n_2}
= \frac{1}{(\sqrt{2}\beta)^{n_1+n_2}}
 \{\ve + \hy\}_{n_1+n_2} \\
%%%%%%%%%%%%%%%%%%
= \frac{1}{(2 \beta)^{n_1+n_2}} \sum_{n_3=0}^{n_1+n_2}
\binom{n_1+n_2}{n_3}
 \{ C_{n_1+n_2-n_3} (\ve) \otimes C_{n_3} (\hy)\}_{n_1+n_2}.
\end{multline}
Finally, we have to substitute the above equations (\ref{eq:m-4}),
(\ref{eq:m-5}) and (\ref{eq:m-6}) into eq.~(\ref{eq:m-3}).
As a result, we arrive at the identity
\begin{multline}
  \label{eq:m-7}
\frac{1}{q^2} C_{nlm} (\hx) 
= \frac{(-1)^n}{p^2}
\sum_{n_1=0}^n \binom{n}{n_1} (-u)^{n_1}
 \sum_{n_2=0}^\infty
v^{n_2}
\binom{n_1 + n_2+1}{n_2}   \sum_{n_3=0}^{n_1+n_2} \binom{n_1+n_2}{n_3} \\
%%%%%%%%%%%%%%
\times
\{ C_{n-n_1} (\ve) \otimes \{ C_{n_2} (\hk) \otimes
 \{ C_{n_1+n_2-n_3} (\ve) \otimes C_{n_3}(\hy) \}_{n_1+n_2} \}_{n_1}
 \}_{nlm},
\end{multline}
where the short-hand notations have been introduced
\begin{equation}
  \label{eq:def-u-v}
  u = \frac{\alpha}{\beta}, \quad
v = \frac{k}{2\beta} = \frac{\sqrt{(\beta-\alpha)^2 + \vvk^2}}{2\beta}.
\end{equation}
The above equation (\ref{eq:m-7}) gives the connection between the
hyperspherical harmonics $C_{n}(\hx)$ and $C_{n_3}(\hy)$.
As is seen, among three summations only one (over the index $n_2$) is
infinite.
Expansion (\ref{eq:m-7}) is the farthest point one can reach using the
angular momentum technique.

If we substitute the expansion (\ref{eq:m-7}) into the integral
(\ref{eq:f-fact-1}), then, due to the orthogonality of HSH, only the terms
with $n_3=n$ will remain.
Thus, the matrix element will be defined by the two summations one of which
is infinite.
Below we demonstrate that the infinite summation can be evaluated in closed
form as the derivatives of a some elementary function.

%%%%%%%%%%%%%%%%%%%%%%%%%

\subsection{Explicit expressions for the coefficients of the multipole
  expansion}
\label{sec:calc-mult-coeff}

 The tensor product in the expansion (\ref{eq:m-7}) has a remarkable
property.
Namely, the internal, as well as external, tensor products are ``minimal'',
i.e. the rank of the product is equal to the sum of ranks of HSH
entering the product.
The intermediate product having the rank $n_1$, contains tensor multipliers
with the ranks $n_2$ and $n_1+n_2$.
Such tensor product can be considered as ``maximal'' since its rank is equal
to the difference of the ranks of the constituent tensors.
It turns out that the four-dimensional Clebsch-Gordan coefficients
corresponding to the minimal or maximal tensor products can be evaluated in the
closed form, see \cite{avm05:_o4_jpa}.
As a consequence, the tensor construction in eq.~(\ref{eq:m-7}) can be
significantly simplified.

The calculation of the three-fold tensor product in eq.~(\ref{eq:m-7})
simplifies also by the fact that the hyperspherical harmonics of the unit basis
vector $\ve$ reduces to a constant number \cite{avm05:_o4_jpa},
\begin{equation}
  \label{eq:m-8}
  C_{n l m} (\ve) = \sqrt{n+1} \, \delta_{l,0} \, \delta_{m,0}.
\end{equation}
Using this equation and the definition of the tensor product of HSH
\cite{avm05:_o4_jpa}, we can write the tensor product from the right-hand side
of eq.~(\ref{eq:m-7}) as 
\begin{multline}
  \label{eq:m-9}
T^{(n_1 n_2 n_3)}_{nlm} \equiv
\{ C_{n-n_1} (\ve) \otimes \{ C_{n_2} (\hk) \otimes
 \{ C_{n_1+n_2-n_3} (\ve) \otimes C_{n_3}(\hy) \}_{n_1+n_2} \}_{n_1} \}_{n lm}
 \\
%%%%%%%%%%%%%%%%%%%
= \sum_{l_2, l_3} \sum_{m_2,m_3}
  C^{nlm}_{(n-n_1) 00;\, n_1 lm} \,
C^{n_1 lm}_{n_2 l_2 m_2;\, (n_1+n_2) l_3 m_3}
C^{(n_1+n_2) l_3 m_3}_{(n_1+n_2-n_3) 00; \, n_3 l_3 m_3} \\
%%%%%%%%%%%%%%%
\times
\sqrt{(n-n_1+1)\, (n_1+n_2-n_3+1)} 
C_{n_2 l_2 m_2} (\hk) C_{n_3 l_3 m_3} (\hy),
\end{multline}
where $C^{nlm}_{n' l'm';\, n'' l''m''}$ denotes the Clebsch-Gordan
coefficients of the $O(4)$ group.
%% As was noted above, the Clebsch-Gordan coefficients in the above tensor
%% product can be written in closed form \cite{avm05:_o4_jpa}.

We have to calculate the summation over the index $n_2$ in (\ref{eq:m-7}).
First, we note the explicit expression for HSH $C_{n_2 l_2 m_2} (\hk)$,
\begin{equation}
  \label{eq:ap2-1}
  C_{n_2 l_2 m_2} (\hk)
= (-i)^{l_2} \sqrt\frac{2l_2+1}{n_2+1} \chi^{n_2/2}_{l_2} (2\theta_0)
 C_{l_2 m_2} (\theta,\phi),
\end{equation}
where $C_{l_2 m_2} (\theta,\phi)$ are conventional spherical harmonics
depending on the spherical angles $\theta,\phi$ of the vector $\vvk$.
The hyper-angle $\theta_0$ is defined by
\begin{equation}
  \label{eq:theta0}
  \cos\theta_0 = \frac{\beta-\alpha}{k} =
 \frac{\beta-\alpha}{\sqrt{(\beta-\alpha)^2 + \vvk^2}}, \quad
\sin\theta_0= \frac{|\vvk|}{\sqrt{(\beta-\alpha)^2 + \vvk^2}}.
\end{equation}
The functions $\chi^{n_2/2}_{l_2}$ in eq.~(\ref{eq:ap2-1}) are the generalized
characters of the $O(3)$ rotation group which are connected to the Gegenbauer
polynomials \cite{Varsh}.
Thus, the explicit expression for HSH $C_{n_2 l_2 m_2}$ has the form,
\begin{equation}
  \label{eq:ap2-2}
  C_{n_2 l_2 m_2} (\hk)  
= (-i)^{l_2} \, (2l_2)!!\, \sqrt\frac{(2l_2+1)\,(n_2-l_2)!}{(n_2+l_2+1)!}
 (\sin\theta_0)^{l_2} C^{l_2+1}_{n_2-l_2} (\cos\theta_0)
 C_{l_2 m_2} (\theta,\phi),
\end{equation}
where $C^{l_2+1}_{n_2-l_2}$ is the Gegenbauer polynomial.
Substituting this equation and eq.~(\ref{eq:m-9}) into eq.~(\ref{eq:m-7}) 
we arrive at the multipole expansion
\begin{equation}
  \label{eq:decomp-2}
  \frac{1}{q^2} C_{nlm} (\hx)
= \frac{(-1)^n}{p^2}  \sum_{n_3=0}^\infty \sum_{l_3=0}^{n_3}
\sum_{l_2} 
A_{n,l;\, n_3,l_3}^{(l_2)}
C^{l_3 0}_{l 0\, l_2 0} 
\sum_{m_2, m_3} C^{lm}_{l_2 m_2\, l_3 m_3} 
C_{l_2 m_2} (\theta,\phi) C_{n_3 l_3 m_3} (\hy),
\end{equation}
where $C^{lm}_{l_2 m_2\, l_3 m_3}$ is the three-dimensional Clebsch-Gordan
coefficient, and the index $l_2$ runs from $|l-l_3|$ to $l+l_3$ so that
the sum $l+l_2+l_3$ is an even number.
This is due to the fact that the coefficients $C^{l_3 0}_{l 0\, l_2 0}$ vanish
at odd values of the sum $l+l_2+l_3$.
The coefficients $A$ in the above equation are defined by
\begin{multline}
  \label{eq:def-a}
A_{n,l;\, n_3,l_3}^{(l_2)}
= (2 i  \sin\theta_0)^{l_2} \, l_2! \, (2l_2+1) 
\sqrt\frac{(n+l+1)!}{(n-l)!\, (n_3+l_3+1)!\, (n_3-l_3)!}\, \\
%%%%%%%%%%%%%%%%%
\times
\sum_{n_1} \binom{n-l}{n-n_1}
\frac{ (-u)^{n_1}}{(n_1+l+1)!}
 \sum_{n_2} v^{n_2}
C^{l_2+1}_{n_2-l_2} (\cos\theta_0)\,
 \frac{(n_1+n_2+l_3+1)!\, (n_1+n_2-l_3)!}{(n_2+l_2+1)!\,(n_1+n_2-n_3)!},
\end{multline}
where the summations are performed over all non-negative values of indices
$n_1$ and $n_2$ so that all binomial coefficients and factorials remain
finite.
As is seen, the summation over $n_1$ is finite while the summation over $n_2$
is infinite.
These summations are calculated in Appendix~\ref{sec:calculation-sum-over}.
The sum over $l_2$ in (\ref{eq:decomp-2}) has simple physical meaning.
It is the multipole expansion with respect to the ``momentum transfer''
vector $\vvk$.

As was pointed above, the multipole expansion for the product 
$q^{-4} C_n (\hx)$ can be derived from the expansion for the function
$q^{-2} C_n (\hx)$ by applying eq.~(\ref{eq:xi-7}).
Omitting details of some routine transformations, we present the final result,
\begin{multline}
  \label{eq:decomp-q-4}
  \frac{1}{q^4} C_{nlm} (\hx)
= \frac{(-1)^n}{2 \alpha^2 p^2}  
 \sum_{n_3=0}^\infty \sum_{l_3=0}^{n_3}
 \sum_{l_2} 
(-1)^{n_3} (-i)^{l_2}
B_{n,l;\, n_3,l_3}^{(l_2)}
C^{l_3 0}_{l 0\, l_2 0} \\
%%%%%%%%%%
\times
\sum_{m_2, m_3} C^{lm}_{l_2 m_2\, l_3 m_3} 
C_{l_2 m_2} (\theta,\phi) C_{n_3 l_3 m_3} (\hy),  
\end{multline}
where the coefficients $B_{n,l;\, n_3,l_3}^{(l_2)} (\theta_0)$ are
\begin{equation}
  \label{eq:def-b}
B_{n,l;\, n_3,l_3}^{(l_2)}
= - \left(\frac{|\vvk|}{\beta}\right)^{l_2} 
 \frac{l_2! \, (2l_2+1)}{2(n+1)}
\sqrt\frac{(n+l+1)!\,(n_3+l_3+1)!}{(n-l)!\, (n_3-l_3)!}\,
P^{(l_2)}_{n,l;\, n_3, l_3},
\end{equation}
and the functions $P^{(l_2)}_{n,l;\, n_3, l_3}$ are defined by
eq.~(\ref{eq:sum-3a}) of Appendix~\ref{sec:calculation-sum-over}.

At zero ``momentum transfer'' (i.e. at $\vvk=0$), the properties of the
three-dimensional Clebsch-Gordan coefficients in the expansion
(\ref{eq:decomp-q-4}) lead to the vanishing of all terms  with $l_2 > 0$ and
$l \neq l_3$.
Thus, the expansion (\ref{eq:decomp-q-4}) reduces to
\begin{equation}
  \label{eq:decomp-2-dip}
  \frac{1}{q^4} C_{nlm} (\hx)
= \frac{1}{2 \alpha^2 p^2} \sum_{n_3=l}^\infty (-1)^{n+n_3}
B_{n,l;\, n_3,l}^{(0)} C_{n_3 l m} (\hy),
\end{equation}
where the parameter $B^{(0)}_{n, 0; n_3, 0} (\vvk=0)$ is calculated in
Appendix~\ref{sec:expl-P-params}.

For the sake of completeness we note that the multipole expansion
(\ref{eq:m-7}) can also be written as the series containing the
four-dimensional solid harmonics,
${\cal C}_{n_2 l_2 m_2} (\vk) = k^{n_2} C_{n_2 l_2 m_2} (\hk)$,
\begin{equation}
  \label{eq:m-10}
  \frac{1}{q^2} C_{nlm} (\hx)
= \frac{(-1)^n}{p^2} \sum_{n_2,n_3,l_2,l_3} 
C^{l_3 0}_{l 0\, l_2 0} C^{lm}_{l_2 m_2\, l_3 m_3} 
C_{n_3 l_3 m_3} (\hy) \, {\cal C}_{n_2 l_2 m_2} (\vk) \,
D_{n_2 n_3,l l_3} (\alpha,\beta),
\end{equation}
where 
%% ${\cal C}_{n_2 l_2 m_2} (\vk)$ are  and
the coefficients $D_{n_2 n_3,l,l_3}(\alpha,\beta)$ are defined by
\begin{equation}
  \label{eq:m-11}
  \begin{split}
  D_{n_2 n_3,l l_3} (\alpha,\beta) =& D
\sum_{n_1} 
 \frac{(n_1+n_2+l_3+1)!\, (n_1+n_2-l_3)!}{(n-n_1)!\,(n_1+n_2-n_3)!\,
(n_1+l+1)!\, (n_1-l)!}
(\alpha/\beta)^{n_1}, \\
%%%%%%%%%%%%%%%%%%%%%%%%%%%%%
D =& (-1)^{l_2} (2\beta)^{-n_2} 
\sqrt\frac{(n+l+1)!\, (n-l)! (2l_2+1)}{(n_3+l_3+1)!\, (n_3-l_3)!\,
(n_2+l_2+1)!\, (n_2-l_2)!} 
  \end{split}
\end{equation}
The summation over $n_1$ in eq.~(\ref{eq:m-11}) can be calculated in closed
form which yields the hypergeometric function ${}\!_3F_2$, see
ref.~\cite{Bateman-I}.
%% (\textbf{Unfortunately, $D$-coefs diverge on the bound-free transitions!})
Thus, the expression for the coefficients $D_{n_2 n_3, ll_3}$ becomes
\begin{multline}
  \label{eq:m-12}
  D_{n_2 n_3, ll_3} (\alpha,\beta) = D\,
\frac{(n_2+l+l_3+1)!\, (n_2+l-l_3)!}{(n_2-n_3+l)!\,(n-l)!\, (2l+1)!} \\
%%%%%%%%%%%%%
\times
\left( \frac{\alpha}{\beta} \right)^l\, 
{}_3F_2 \left(
    \begin{array}{cccc}
      l-n, & n_2+l-l_3+1, & n_2+l+l_3+2; & - \frac{\alpha}{\beta} \\
      & n_2-n_3+l+1, & 2l+2 &
    \end{array}
 \right).
\end{multline}
The analysis of the series (\ref{eq:m-10}) shows that the region of
convergence with respect to the index $n_2$ is defined by the two conditions,
$|\alpha| < |\beta|$ and $|(\alpha-\beta)^2+\vvk^2| < 2 |\beta|$.
The series in (\ref{eq:m-10}) diverges when one of parameters
$\alpha$, $\beta$ is purely imaginary and another one is real.

The convergence of the above expansions (\ref{eq:decomp-q-4}) and
(\ref{eq:m-10}) with respect to the index $n_3$ of HSH is practically not
important because these expansions in $n_3$ will always be truncated after 
the integration over the hyper-angles of the unit vector $\hy$.

%%%%%%%%%%%%%%%%%%%%%%%%%%%%%%%%

\section{Properties of the matrix elements}
\label{sec:matrix-elements}

In this section we derive the explicit expressions for the matrix element in
eq.~(\ref{eq:f-fact-1}) in terms of the coefficients $B^{(l_2)}_{n,l;n_3,l_3}$
and $P^{(l_2)}_{n,l;n_3,l_3}$ introduced in the previous section. 
In sec.~\ref{sec:mel-general} the general formulas are given and few examples
are considered.
In sec.~\ref{sec:dipole-limit} we analyze the long-wave (i.e. dipole) limit of
the matrix elements.

%%%%%%%%%%%%%%%%%%%%%%%%%

\subsection{Matrix elements in general case}
\label{sec:mel-general}

In order to derive an expression for the form-factor matrix element,
we replace in the decomposition (\ref{eq:decomp-q-4}) HSH $C_{nlm}$ with the
normalized HSH $Y_{nlm}$ and insert the resulting expression into
(\ref{eq:f-fact-1}).
This leads to the equation
\begin{multline}
  \label{eq:me-int-2}
  F_{n'l',nl} = \sum_{n_3,l_3,l_2} 
i^{l_2} \sqrt\frac{\alpha (n+1)}{\beta (n_3+1)}
 B^{(l_2)}_{n,l;n_3,l_3} 
C^{l_3 0}_{l 0\, l_2 0} \\
%%%%%%%%%%%%%
\times
\sum_{m_2, m_3} C^{lm}_{l_2 m_2\, l_3 m_3} 
C_{l_2 m_2} (\theta,\phi)
\int Y^*_{n'l'm'} (\hy) Y_{n_3 l_3 m_3} (\hy) \,d \Omega_\vy.
\end{multline}
Since HSH are orthonormal, the integral over the hypersphere is equal to unity
for $n_3 l_3 m_3=n'l'm'$ and is zero otherwise.
As a consequence, we obtain the expression
\begin{equation}
  \label{eq:me-int3}
  F_{n'l',nl} = \sqrt\frac{\alpha (n+1)}{\beta (n'+1)}
\sum_{l_2=|l-l'|}^{l+l'} i^{l_2} 
C^{l' 0}_{l 0\, l_2 0}
C^{lm}_{l_2 m_2\, l' m'} C_{l_2 m_2} (\theta,\phi)\,
B^{(l_2)}_{n,l;n',l'}.
\end{equation}
If the transitions between the states of the hydrogen atom are considered then
the factor under the square root reduces to unity as follows from the
definition (\ref{eq:mel-5}) of parameters $\alpha$ and $\beta$.

The above formula for the form-factors $F_{n'l',nl}$ can be re-written in
terms of radial integrals.
Using explicit expressions for the radial hydrogen wave functions and
decomposing the plane-wave in eq.~(\ref{eq:form-fact-1}) over the
three-dimensional spherical functions, eq.~(\ref{eq:me-int3}) can be brought
to the form
\begin{multline}
  \label{eq:I-F11}
 \int_0^\infty r^{l+l'+2} e^{- (\alpha+\beta)\,r}\, 
j_{l_2} \left( |\vvk| r\right)
\,\Phi(l'-n',2l'+2; 2 \beta r) \, \Phi(l-n,2l+2;2\alpha r) \, dr \\
%%%%%
= (-1)^{l+l_2+l'+1} \,
\frac{l_2!\, (2l+1)!\, (2l'+1)!}{ (2 \alpha)^{l+1} (2 \beta)^{l'+2}} 
\left( \frac{|\vvk|}{\beta} \right)^{l_2} P^{(l_2)}_{n,l;\,n',l'} (u,w),
\end{multline}
where the parameter $w$ is defined by eq.~(\ref{eq:geg-arg}) of
Appendix~\ref{sec:calculation-sum-over}, $\Phi (l-n,2l+2; 2 \alpha r)$ is
the confluent hypergeometric function and $j_{l_2}$ is the spherical Bessel
function.
The definition (\ref{eq:sum-3a}) of functions $P^{(l_2)}_{n,l;\,n',l'}$
implies the following limitations on the indices in eq.~(\ref{eq:I-F11}),
\begin{equation}
  \label{eq:index_limits}
  l \le n, \quad l' \le n', \quad l-l_2+l'+1 \ge 0.
\end{equation}
(The requirement for the combination $l+l_2+l'$ to be an even number is not
necessary in eq.~(\ref{eq:I-F11}).)
Thus, we have shown that the integrals of the kind
(\ref{eq:I-F11}) can be written as a three-fold derivative of an elementary
function (cf. eqs.~(\ref{eq:I-F11}) and (\ref{eq:sum-3a})).
It turns out that for small values of indices $n$ and $l$ (or $n'$ and $l'$)
the functions $P^{(l_2)}_{n,l;\,n',l'}$ can be written in a compact form as a
combination of Gegenbauer polynomials.

\begin{figure}[htbp]
  \centering
  \includegraphics[width=8cm]{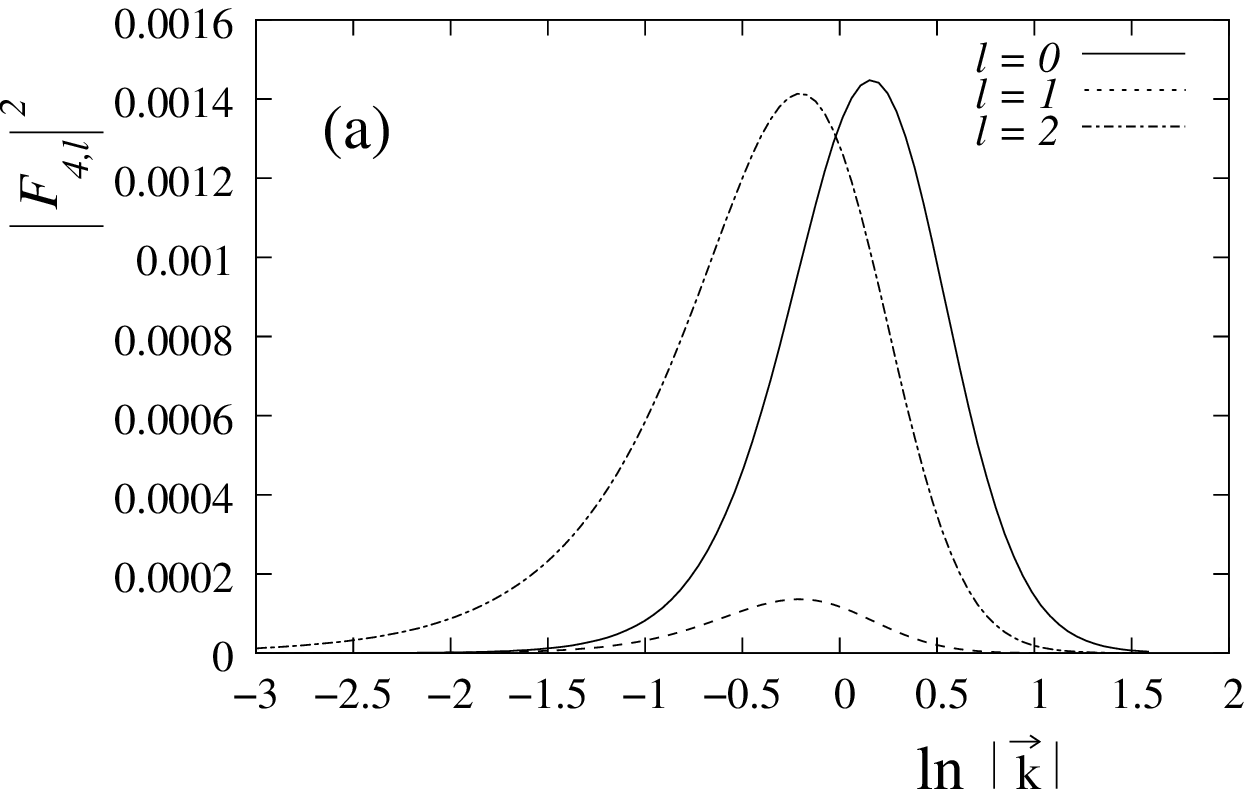} \;
  \includegraphics[width=7.5cm]{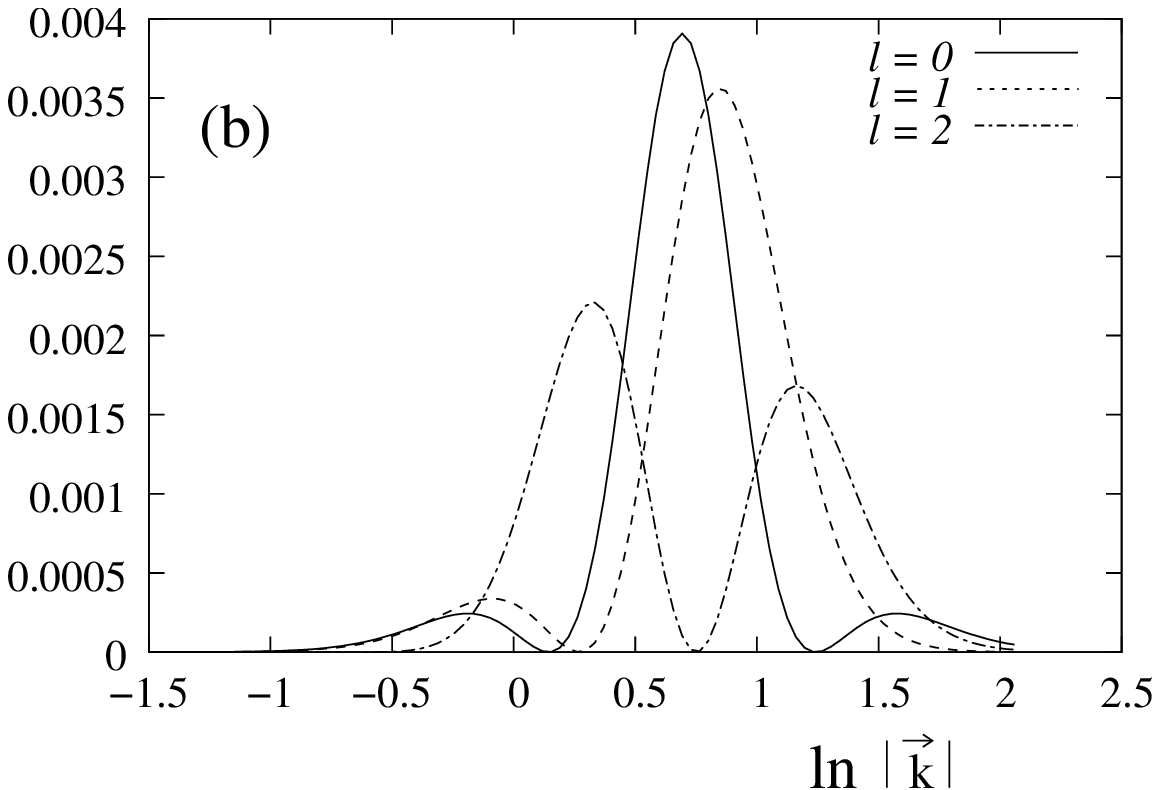}
  \caption{The parameters $| F_{4,l}|^2$. (a) the matrix elements for the
    hydrogen atom: $\alpha=1$, $\beta=1/5$.
    (b) the matrix elements for Sturmians: $\alpha=\beta=1$.}
  \label{fig:b-s}
\end{figure}

For example, the matrix elements for the transitions from the ground state
$n=l=0$ can be written as
\begin{equation}
  \label{eq:1s-1}
F_{n,l} = \int_0^\infty R_{0,0} (\alpha,r) j_l (kr) R_{n,l} (\beta,r)\, r^2 dr
= - \frac{l !}{2} \left(\frac{k}{\beta} \right)^{l}
\sqrt\frac{\alpha (n+l+1)!}{\beta (n+1)\,(n-l)!}\,
P^{(l)}_{0,0;\,n,l},
\end{equation}
where $R_{n,l}$ are normalized radial wave functions and the function
$P^{(l)}_{0,0;\,n,l}$ is the combination of two Gegenbauer polynomials
(see eq.~(\ref{eq:s-4}) of Appendix~\ref{sec:p-limiting-cases}).
We remind that the index $n$ above is equal to the principal quantum number
minus unity.
For hydrogen matrix elements one has $\alpha=1$ and $\beta=1/(n+1)$ while
for Sturmian matrix elements the parameters $\alpha$ and $\beta$ can be
arbitrary numbers.
As an example, on fig.~\ref{fig:b-s} the parameters $|F_{n,l}|^2$
corresponding to transition $1s \to 5l$ are plotted as functions of 
$\ln |\vvk|$. 
(The reason for such choice of abscissa variable is explained in
\cite{miller_platzman57:_inelast_collisions_he_GOS,inokuti71:_GOS_Bethe_inelastic}.)
As is seen, the hydrogen matrix elements have strong maxima at some values of
$|\vvk|$.
The Sturmian matrix elements behave in less regular manner, e.g. they can have
several peaks at different values of the momentum $|\vvk|$.

%%%%%%%%%%%%%%%%%%%%%%%%%%%%%%%%%%%%%%

\subsection{The dipole limit}
\label{sec:dipole-limit}

In the dipole limit we have that $\vvk \to 0$ and, hence, it is necessary to
retain only the terms of zeroth- and first-order with respect to $|\vvk|$.
Such terms correspond to the two values of the multipole index, $l_2=0$ and
$l_2=1$.
The derivation of the parameters $P^{(l_2)}_{n,l;\,n',l'}$ for such values
of $l_2$ is considered in Appendix~\ref{sec:p-limiting-cases}.

According to eq.~(\ref{eq:dip-1b}) of Appendix~\ref{sec:p-limiting-cases},
the parameters with $l_2=0$ can be written in closed form in terms of Gauss
hypergeometric function
\begin{multline}
  \label{eq:dip-4}
  P^{(0)}_{n,l;\, n',l} (u) = \frac{4 (-1)^{n'}}{(2l+1)!}
\left( \frac{-4u}{(1-u)^2} \right)^{l+1}
\left(\frac{1-u}{1+u} \right)^{n+n'+1}
\frac{n'+1-(n+1)u}{(1+u)^2}\\
%%%%%%%%%%%%%%%%%%%%%
\times \,
{}_2F_1 \left(l-n, l-n'; 2l+2; \frac{-4u}{(1-u)^2} \right).
\end{multline}
The Gauss function in this equation is a polynomial whose order is 
$\min (n-l,n'-l)$.
It is easy to see that at $u=(n'+1)/(n+1)$ (as well as at $u=1$) the
parameter $P^{(0)}_{n,l;\, n',l}$ is
\begin{equation}
  \label{eq:p0-delta}
 P^{(0)}_{n,l;\, n',l} = - 2 (n+1) \frac{(n-l)!}{(n+l+1)!} \delta_{n,n'}.
\end{equation}
and the form-factor matrix element becomes
\begin{equation}
  \label{eq:form-f-delta}
  F_{n'l',nl} \Bigr|_{\vvk=0} = \delta_{n,n'}.
\end{equation}
This reflects the orthogonality property of the hydrogen wave functions (or,
at $u=1$, Sturmian functions from the same set).

The parameter $P^{(1)}_{n,l;n',l-1}$ describes the dipole transitions
and expresses as (see eq.~(\ref{eq:dip-3}) below)
\begin{multline}
  \label{eq:dip-5}
P^{(1)}_{n,l;\,n',l-1} (u) =  \frac{2 (-1)^{n'}}{3\, (2l+1)!} 
\left( \frac{-4u}{(1-u)^2} \right)^{l+1}\,
\left( \frac{1-u}{1+u} \right)^{n+n'} 
\Bigl\{ (n'+l+1) (n'+l+2) f_0 \\
%%%%%%%%%%%%%%%
+ 2 (n'+l+1) ( n'-l+1) f_1 + (n'-l+1) (n'-l) f_2 \Bigr\},
\end{multline}
where the coefficients $f_k$ are defined by
\begin{equation}
  \label{eq:def-f}
  f_k = \left( \frac{1-u}{1+u} \right)^{2-k}
\frac{(n+1)u - n'-2+k}{(1+u)^2}\,
{}_2F_1 \left(l-n, l-1-n'+k; 2l+2; \frac{-4u}{(1-u)^2} \right).
\end{equation}
In the case of matrix elements between the hydrogen states, we have that
$u=(n'+1)/(n+1)$ and the coefficient $f_1$ vanishes.
Thus, the matrix element in this case is a combination of two Gauss
hypergeometric functions in agreement with the well-known Gordon formula
\cite{gordon29}.
The expression for the coefficient $P^{(l_2)}_{n,l;n',l+1}$ can be derived
from eq.~(\ref{eq:dip-5}) using the symmetry property (\ref{eq:p-symm-1})
which in the dipole case has the form
\begin{equation}
  \label{eq:symm-dip}
    P^{(1)}_{n,l;\,n',l'} (u)
= u^{n+n'+2 l'+2} P^{(1)}_{n',l';\,n,l} (1/u).
\end{equation}

%%%%%%%%%%%%%%%%%%%%%%%%%%%%

\section{Conclusion}
\label{sec:conclusion}

In the presented paper we have applied the method of Fock's stereographic
projections to the calculation of matrix elements involving the
hydrogen-type wave functions.
We have shown (sec.~\ref{sec:ster-proj}) that the problem of the calculation
of matrix elements is equivalent to the derivation of a multipole expansion of
the four-dimensional hyperspherical harmonics with their arguments
defined by means of the shifted stereographic projection.

There is a conceptual similarity between our approach and the work
\cite{manakov_jetp01:_free_param_cfg}.
In that paper the two-photon transitions between the states of the hydrogen
atom have been considered using the representation of the Coulomb-Green
function with two free parameters.
Our multipole expansions of the hydrogen wave function in momentum space (see
eqs.~(\ref{eq:decomp-2}) and (\ref{eq:decomp-q-4})) also contain free
parameters $\beta$ and $\vvk$.
By ``tuning'' the energy parameter $\beta$, we can move the argument of HSH to
any desirable point on the hypersphere.
Usually, the choice of the parameter $\beta$ is fixed by demanding the
orthogonality of HSH entering the integral.
The same approach was used in \cite{manakov_jetp01:_free_param_cfg}.
The proper choice of free parameters can result in the significant
simplification of expressions for the compound matrix elements or in the
imporoving the accuracy of numerical computations.
The method developed in the presented paper can also be used to generalize the
results of
\cite{manakov_jetp01:_free_param_cfg,manakov_jetp04:_cfg_hyperpolarizability}
in order to account the non-dipole effects.

The expression for the form-factor matrix element in terms of the multipole
functions $P^{(l_2)}_{n,l;\,n',l'}$ is given in
sec.~\ref{sec:matrix-elements}.
We showed that the functions $P^{(l_2)}_{n,l;\,n',l'}$ are, in fact, integrals
between the radial hydrogen wave functions and the spherical 
Bessel function arising from the multipole expansion of the boost operator
$\exp(i \vvk\cdot\var)$.

It turns out, that the closed expression for $P^{(l_2)}_{n,l;\,n',l'}$ in
terms of special function of a single variable does not exist in general case
of arbitrary values of the indices $n,l,n',l',l_2$ and arguments
$\alpha,\beta$ and $\vvk$.
 Only in some particular situations, e.g.~in the long-wave (i.e. dipole) limit
as well as for the transitions from the ground $1s$-state, the functions
$P^{(l_2)}_{n,l;\,n',l'}$ can be written as a combination of Gauss
hypergeometric functions\footnote{more precisely, as Jacobi or Gegenbauer
  polynomials, see Appendix \ref{sec:p-limiting-cases}}.
Surprisingly, there exists a remarkably simple differential representation
for coefficients $P^{(l_2)}_{n,l;\,n',l'}$. 
It is given by eq.~(\ref{eq:sum-3a}) of
Appendix~\ref{sec:calculation-sum-over}.
This differential formula can serve as a source for many explicit expressions
for functions $P^{(l_2)}_{n,l;\,n',l'}$.
The differential representation can also be effectively used for the
derivations of the recurrence relations for $P$-functions.
A number of such relations is derived in Appendix~\ref{sec:recurrences}.
We note also that recursions involving radial hydrogen matrix elements of the
powers $r^k$ have been considered in \cite{hey06:h_matr_elem}.

The results of the presented paper can also be generalized on the case of
bound-free transitions.
In general, this can be achieved by applying the replacements
$n \to \pm i/\sqrt{E}$, where $E>0$ is the energy of the continuum state.
At this stage the differential operators in eq.~(\ref{eq:sum-3a}) become
undefined.
However, this problem can be avoided by using the contour representations for
the derivatives.

The matrix elements of the operator of electromagnetic interaction 
${\vec \epsilon}\, \exp(i \vvk \cdot \var)$ can also be
derived using the formalism developed in this paper.
In this case one should replace $q^{-4}$ with $(\bm\epsilon \cdot \hx) q^{-4}$
in eq.~(\ref{eq:f-fact-1}) where $\bm\epsilon=(0, {\vec \epsilon})$ is the
four-dimensional photon polarization vector.
Then, one can apply the Wigner-Eckart theorem to the ensuing integral.
The corresponding results can be written again in terms of the functions
$P^{(l_2)}_{n,l;\,n',l'}$.

One of the possible applications of the technique developed above could be the
calculation of the two-center Coulomb integrals.
Namely, by applying the Fourier transform to the function 
$1/| {\vec r}_1 - {\vec r}_2|$ one arrives at the form-factor integrals for
which the differential representations can be used.
The work on this problem is in progress.
%% Another possible application would be the analysis of the three-body problem
%% in momentum space.
%% We note There are 
%% The problem of the helium atom and the hydrogen molecule in momentum space 
%% were also studied e.g. in
%% \cite{mcweeny_coulson49:_helium_momentum_space,mcweeny49:_h_molecule_momentum_space}.

\acknowledgements
This work has been supported in part by CRDF and RF ministry of education
in BRHE program under the Grant No Y2-CP-10-02 (AVM).
%% AVM also thanks Max-Planck Insitute for the Physics of Complex Systems
%% (Dresden) for hospitality.

\appendix

%%%%%%%%%%%%%%%%%%%%%%%%%%%

\section{The calculation of summations in eq.~(\ref{eq:def-a}).}
\label{sec:calculation-sum-over}

In this Appendix we calculate the two summations in (\ref{eq:def-a}).
We begin by calculating the sum over $n_2$ which we denote as $S_{n_1}$,
\begin{equation}
  \label{eq:ap2-3}
S_{n_1} 
=  \sum_{n_2=l_2}^\infty v^{n_2} C^{l_2+1}_{n_2-l_2} (\cos\theta_0)
\frac{(n_1+n_2+l_3+1)!\,(n_1+n_2-l_3)!}{(n_1+n_2-n_3)!\,(n_2+l_2+1)!}.
\end{equation}
Replacing the summation variable by $k=n_2-l_2$ we obtain
\begin{equation}
  \label{eq:ap2-5}
S_{n_1} 
=  v^{l_2} \sum_{k=0}^\infty v^k C^{l_2+1}_k (\cos\theta_0)
\frac{(k+n_1+l_2+l_3+1)!\,(k+n_1+l_2-l_3)!}{(k+n_1-n_3+l_2)!\,(k+2 l_2+1)!}.
\end{equation}
Factorials in this identity can be removed by introducing the differential
operators
\begin{multline}
  \label{eq:sum-1}
\frac{(k+n_1+l_2+l_3+1)!\,(k+n_1+l_2-l_3)!}{(k+n_1-n_3+l_2)!\,(k+2 l_2+1)!}\\
%%%%%%%%%%%%
= (-1)^{n_1+n_3+l_2} 
\deriv_t^{n_3-l_3}
t^{n_3+l_3+1} 
\deriv^{n_1-l_2+l_3}_\tau
(t+\tau)^{-(k+2l_2+2)} \biggr|_{t=1, \tau=0},
\end{multline}
where $\partial_t = \partial/\partial t$ and
$\partial_\tau= \partial/\partial \tau$.
Inserting equation (\ref{eq:sum-1}) into (\ref{eq:ap2-5}) one arrives
at the series which can be evaluated using the generating function for
Gegenbauer polynomials \cite{Bateman-II},
\begin{multline}
  \label{eq:sum-2}
(t+\tau)^{-(2l_2+2)} \sum_{k=0}^\infty 
\left(\frac{v}{t+\tau} \right)^k C_k^{l_2+1}(\cos \theta_0)
= (t+\tau)^{-(2l_2+2)} 
\left(1-\frac{2v\cos\theta_0}{t+\tau} + \frac{v^2}{(t+\tau)^2}
\right)^{-l_2-1} \\
%%%%%%%%%%%
= [v^2 - 2 v \cos\theta_0 (t+\tau) + (t+\tau)^2 ]^{-l_2-1}.
\end{multline}
Thus, the parameter $S_{n_1}$ can be written as
\begin{equation}
  \label{eq:sn1}
  S_{n_1} = (-1)^{n_1+n_3+l_2} \, v^{l_2}
\deriv^{n_3-l_3}_t
t^{n_3+l_3+1} \,
\deriv^{n_1-l_2+l_3}_\tau
[v^2 - 2 v \cos\theta_0 (t+\tau) + (t+\tau)^2]^{-l_2-1} 
 \biggr|_{t=1, \tau=0}.
\end{equation}
Now we have to evaluate the sum over $n_1$ in eq.~(\ref{eq:def-b}),
\begin{equation}
  \label{eq:sum-n1-1}
  S_n = \sum_{n_1=l}^n \binom{n-l}{n-n_1}
\frac{ (-u)^{n_1}}{(n_1+l+1)!} 
 S_{n_1}
= \sum_{n_1=0}^{n-l} \binom{n-l}{n_1}
\frac{ (-u)^{n_1+l}}{(n_1+2l+1)!} S_{n_1+l}.
\end{equation}
Using eq.~(\ref{eq:sn1}) we write $S_n$ in its full form,
\begin{multline}
  \label{eq:sum-n1-2}
  S_n = (-1)^{n_3+l_2} \, v^{l_2}\, \deriv_t^{n_3-l_3} \,
t^{n_3+l_3+1} \sum_{n_1=0}^{n-l} \binom{n-l}{n_1}
\frac{u^{n_1+l}}{(n_1+2l+1)!} \\
%%%%%%%%%%%%%
\times
\deriv_\tau^{n_1+l-l_2+l_3}
[v^2 - 2 v (t+\tau) \cos\theta_0 + (t+\tau)^2]^{-l_2-1} 
\biggl|_{t=1, \tau=0}.
\end{multline}
After some rearrangements using the chain differentiation rule, this identity
can be written in closed form,
\begin{multline}
  \label{eq:sum-3}
S_n =  (-1)^{n_3+l_2}  \frac{v^{l_2}\,u^{-l-1} }{(n+l+1)!}
 \deriv_t^{n_3-l_3}
t^{n_3+l_3+1} 
\deriv_\tau^{n-l} 
(u+ \tau )^{n+l+1} \deriv^{l-l_2+l_3}_\tau \\
%%%%%%%%%%%%%%
\times
\left[ v^2 - 2 v (t+\tau)  \cos\theta_0 + (t+\tau)^2 \right]^{-l_2-1} 
\biggl|_{t=1, \tau=0}.
\end{multline}
As is seen, the two-fold summation in (\ref{eq:def-a}) reduces to the
derivatives of an elementary function.

According to eq.~(\ref{eq:xi-7}), in order to calculate the coefficients in the
expansion (\ref{eq:decomp-q-4}) of the function $q^{-4} C_{n} (\hx)$, one has
to evaluate the combination of three functions $S_n$.
Omitting the details of some routine transformations, we present the final
result for the parameters $P^{(l_2)}_{n,l;\,n_3,l_3}$ in (\ref{eq:def-b}),
\begin{multline}
  \label{eq:sum-3a}
P^{(l_2)}_{n,l;\,n_3,l_3} (u,w)
= \frac{u^{-l} }{(n+l+1)!\,(n_3+l_3+1)!}
\deriv_\tau^{n-l} 
(u+ \tau )^{n+l+1}
 \deriv_t^{n_3-l_3}
(1+t)^{n_3+l_3+1} \\
%%%%%%%%%%%%%%
\times
\deriv^{l-l_2+l_3+1}_\tau
\left[ w^2 + (1+u) (t+\tau) + (t+\tau)^2 \right]^{-l_2-1} 
\biggl|_{t, \tau=0},
\end{multline}
where the short-hand notations have been introduced
\begin{equation}
  \label{eq:geg-arg}
 \begin{split}
  w =& \sqrt{1-2v \cos\theta_0 +v^2}
   = \frac{\sqrt{(\alpha+\beta)^2 + \vvk^2}}{2 \beta},
 \end{split}
\end{equation}
We note also the following symmetry property of functions 
$P^{(l_2)}_{n,l;\,n_3,l_3} (u,w)$,
\begin{equation}
  \label{eq:p-symm-1}
  P^{(l_2)}_{n,l;\,n_3,l_3} (u,w)
= u^{n+n_3+2l_3+2} 
P^{(l_2)}_{n_3,l_3;\,n,l} (1/u,w/u).
\end{equation}
This equation can be proved by making the replacements $t \to u \tau$ and
$\tau \to u t$ in the definition (\ref{eq:sum-3a}).

%%%%%%%%%%%%%%%%%%%%%%%%

\section{The explicit expressions for the functions
 $P^{(l_2)}_{n,l;\,n_3,l_3}$}
\label{sec:expl-P-params}

In order to calculate the functions $P^{(l_2)}_{n,l;\,n_3,l_3}$ in
eq.~(\ref{eq:sum-3a}) in an explicit form, we use the generating function for
the Gegenbauer polynomials
\begin{equation}
  \label{eq:d-1}
\left[ w^2 - 2  \xi w (t+\tau) + (t+\tau)^2 \right]^{-l_2-1}
= \sum_{q=0}^\infty w^{-2 l_2-2-q} (t+\tau )^{q} \,
 C^{l_2+1}_{q} \left(- \frac{u+1}{2w} \right).
\end{equation}
Acting by the operator $\deriv^{l-l_2+l_3+1}_\tau$ on this equation one
observes that the terms with $q<l-l_2+l_3+1$ vanish.
Thus, the functions $P^{(l_2)}_{n,l;\,n_3,l_3}$ can be written as a
combination of Gegenbauer polynomials
\begin{equation}
  \label{eq:d-4}
P^{(l_2)}_{n,l;\,n_3,l_3} (u,w) = \frac{u^{l+1}}{w^{l+l_2+l_3+3}} 
\sum_{q =0}^{q_{max}} w^{-q} \, C_q (u)\, C^{l_2+1}_{l-l_2+l_3+q+1}
\left(- \frac{u+1}{2w} \right),
\end{equation}
where $q_{max}=n+n_3-l-l_3$ and parameters $C_q$ are polynomials in $u$,
\begin{equation}
  \label{eq:def-c}
  C_q (u)= \sum_{m=0} \binom{n-l}{m} \binom{n_3-l_3}{q-m} 
\frac{ (l-l_2+l_3+q+1)!}{(2l+1+m)!\, (2l_3+q+1-m)!} u^m,
\end{equation}
where the summation over $m$ is taken over all non-negative integer values at
which factorials remain finite.
The parameters $C_q(u)$ can also be written in the following form
\begin{displaymath}
  C_q (u) = (-1)^q
\frac{(n-l)!\,(n_3-l_3)!\,(l-l_2+l_3+q+1)!}{(n+l+1)!\, (n_3+l_3+1)!\,
q!}\,
\deriv^q_t L^{(2l+1)}_{n-l} (ut)\, L^{(2l_3+1)}_{n_3-l_3} (t) \biggl|_{t=0},
\end{displaymath}
where $L^{(2l+1)}_{n-l}$ is the Laguerre polynomial \cite{Bateman-II}.
This equation can be proved by straightforward calculations.
Thus, the parameter $C_q (u)$ is proportional to the coefficient at the $q$-th
power of the product of two Laguerre polynomials.

One can obtain an explicit expression for $C_q(u)$ in terms of generalized
hypergeometric function ${}_3F_2$.
There are two cases which must be considered separately.
For $q\le n_3-l_3$ the index $m$ in (\ref{eq:def-c}) runs from
$0$ to $\mathrm{min} (q,n-l)$.
This leads to the formula
\begin{equation}
  \label{eq:expr-c-1}
  C_q (u) = \binom{n_3-l_3}{q} \frac{(l-l_2+l_3+q+1)!}{(2l_3+q+1)!\,(2l+1)!}\,
{}_3F_2 \left(
    \begin{array}{rrrr}
       -q, & l-n, & -q-2 l_3-1; & -u \\
      & 2l+2, & n_3-l_3-q+1 &
    \end{array}
 \right).
\end{equation}
For $q > n_3-l_3$ one has that $q-n_3+l_3 \le m \le n-l$.
In this case the above formula is invalid and the correct expression has
the form,
\begin{multline}
  \label{eq:expr-c-2}
  C_q (u) = \binom{n-l}{q-n_3+l_3}
  \frac{(l-l_2+l_3+q+1)!}{(2l-n_3+l_3+q+1)!\,(n_3+l_3+1)!}\,
u^{q-n_3+l_3}\\
%%%%%%%%%%%%%%%%
\times
{}_3F_2 \left(
    \begin{array}{rrrr}
      l+l_3-n-n_3+q, & l_3-n_3, & -l_3-n_3-1 ; & -u \\
      & 2l +2, & l_3-n_3+q+1 &
    \end{array}
 \right).  
\end{multline}
%% We note that the parameter $P^{(l_2)}_{n,l;\,n_3,l_3}$ can be presented in an
%% explicit form as a triple sum.

%%%%%%%%%%%%%%%%%%%%%%

\section{Functions $P^{(l_2)}_{n,l;\,n_3,l_3}$ in particular cases}
\label{sec:p-limiting-cases}

In this appendix we analyze the situations when the functions
$P^{(l_2)}_{n,l;\,n_3,l_3}$ can be written in closed form.
Our consideration is based on the general differential relation
\begin{equation}
  \label{eq:dif-theor}
  \deriv_t^n (\alpha+\beta t)^{n+\gamma} f(t) \bigr|_{t=0}
= \alpha^{2\gamma +n+1} \deriv_t^n (\alpha-\beta t)^{-\gamma-1} \,
 f \left( \frac{\alpha t}{\alpha-\beta t} \right) \biggr|_{t=0}.
\end{equation}
Here, $\alpha,\beta,\gamma$ are arbitrary numbers and $f(t)$ is an arbitrary
function.
This formula can be proved by expanding the function $f$ into Taylor
series and comparing coefficients at the expansion terms.

Let us consider the functions $P$ corresponding to the transition from the
ground $1s$-state.
In this case we have $n=l=0$ and, hence, $l_2=l_3 \equiv l$.
According to the definition eq.~(\ref{eq:sum-3a}), we can write
\begin{equation}
  \label{eq:p-s-1}
P^{(l)}_{0,0;\,n,l} = \frac{u}{(n+l+1)!}
\deriv_t^{n-l} (1+t)^{n+l+1} \deriv_t [w^2 + (u+1)t + t^2]^{-l-1}
\Bigr|_{t=0}.
\end{equation}
In order to calculate the action of the differential operators we
note the formula
\begin{equation}
  \label{eq:s-1}
  (1+t)^{n+l+1} \deriv_t = \deriv_t (1+t)^{n+l+1} - (n+l+1) (1+t)^{n+l}.
\end{equation}
Now the parameter $P^{(l)}_{0,0;n,l}$ can be re-written as
\begin{equation}
  \label{eq:s-2}
  P^{(l)}_{0,0;n,l} = \frac{u}{(n+l+1)!}
\left( \deriv_t^{n-l+1} (1+t) - (n+l+1) \deriv_t^{n-l} \right)
(1+t)^{n+l} [ w^2 + (u+1) t +t^2]^{-l-1} \Bigr|_{t=0}.
\end{equation}
By applying the differential transformation (\ref{eq:dif-theor}) to this
equation we arrive at the equation,
\begin{equation}
  \label{eq:s-3}
  P^{(l)}_{0,0;n,l} = \frac{u}{(n+l+1)!}
\left( \deriv_t^{n-l+1} - (n+l+1) \deriv_t^{n-l} \right) (1-t)
 [ w^2 -2 a t + (tv)^2]^{-l-1} \Bigr|_{t=0},
\end{equation}
where $a=w^2-(u+1)/2$ and the parameter $v$ is defined in 
eq.~(\ref{eq:def-u-v}).
The calculation of derivatives can be performed by expanding the term in
square brackets similarly to eq.~(\ref{eq:d-1}).
After some simple algebraic manipulations we obtain
\begin{multline}
  \label{eq:s-4}
    P^{(l)}_{0,0;n,l} = \frac{(n-l)!}{(n+l+1)!} \frac{u}{w^{2l+3}}
\left( \frac{v}{w} \right)^{n-l}
\biggl( -2 (n+1) ( w- xv) C_{n-l}^{l+1} (x) \\
%%%%%%%%%%%%%%%
+ (n+l+1) \frac{w^2-v^2}{v} C^{l+1}_{n-l-1} (x)
\biggr),
\end{multline}
where the argument of the Gegenbauer polynomial is
\begin{equation}
  \label{eq:def-x}
x= \frac{w^2-(u+1)/2}{v w} = \frac{\alpha^2- \beta^2+\vvk^2}{4 \beta^2 v w}.  
\end{equation}
Similarly, one can derive rather compact expressions for the $P$-coefficients
for the transitions from $2s$ and $2p$ states.
However, for the sake of shortness, we do not present here the corresponding
equations.

\textbf{The dipole limit}\\
Now, let us consider the functions $P^{(l_2)}_{n,l;\, n_3,l_3}$ in the limit
of small $|\vvk|$, i.e. the dipole limit.
In this case only the zeroth- and first-order terms (with respect to
$|\vvk|$) should be taken into account.
Thus, in the dipole limit one has that $w=(u+1)/2$ and
it is necessary to retain only the terms with $l_2=0$ and $l_2=1$ in
eq.~(\ref{eq:sum-3a}).

Let us calculate first the function $P^{(0)}_{n,l;\,n_3,l}$.
Since $\vvk=0$, the term in square brackets in eq.~(\ref{eq:sum-3a}) reduces
to $[w+t+\tau]^{-2}$.
This simplifies the calculations, so that eq.~(\ref{eq:sum-3a}) becomes
\begin{equation}
  \label{eq:dl-1}
 P^{(0)}_{n,l;\,n_3,l} = 
\frac{- u^{-l}\, (2l+2)!}{(n+l+1)!\, (n_3+l+1)!} 
\deriv_\tau^{n-l} 
(u+ \tau )^{n+l+1}
 \deriv_t^{n_3-l}
(1+t)^{n_3+l+1} (w+t+\tau)^{-2l-3} \Bigr|_{t,\tau=0}.
\end{equation}
It is convenient to present the last multiplicand in this equation as
 $(w+t+\tau)^{-2l-3}=-(1/(2l+2)) \deriv_w (w+t+\tau)^{-2l-2}$,
%% Thus, eq.~(\ref{eq:dl-1}) becomes
\begin{equation}
  \label{eq:dl-2}
 P^{(0)}_{n,l;\,n_3,l} = 
\frac{ u^{-l}\, (2l+1)!}{(n+l+1)!\, (n_3+l+1)!} 
\deriv_w
\deriv_\tau^{n-l} 
(u+ \tau )^{n+l+1}
 \deriv_t^{n_3-l}
(1+t)^{n_3+l+1} (w+t+\tau)^{-2l-2} \Bigr|_{t,\tau=0}.  
\end{equation}
Using the auxiliary identity (\ref{eq:dif-theor}) we can calculate the
derivative over $t$ in closed form,
\begin{equation}
  \label{eq:dl-3}
  \deriv_t^{n_3-l}  (w+\tau+t)^{-2l-2} (1+t)^{n_3+l+1} \Bigr|_{t=0}
= \frac{(n_3+l+1)!}{(2l+1)!} (w+\tau)^{-n_3-l-2} (w-1+\tau)^{n_3-l}.
\end{equation}
By applying eq.~(\ref{eq:dif-theor}) to the calculation of the derivative over
$\tau$ one arrives at the Gauss hypergeometric function,
\begin{multline}
  \label{eq:dip-1a}
P^{(0)}_{n,l;\,n_3,l} = \frac{u^{n+1}}{(2l+1)!}
\deriv_w w^{-n-n_3-2} \, (w-1)^{n_3-n}\, z^{l-n}\,
{}_2F_1 (l-n, l-n_3; 2l+2; z),
\end{multline}
where $z=u/[(w-u) (w-1)]$.

The calculation of the derivative over $w$ simplifies by the fact that
$\deriv_w z(w)=0$ at the point $w=(u+1)/2$.
Thus, the parameter $P^{(0)}_{n,l;\,n_3,l}$ assumes the form
\begin{multline}
  \label{eq:dip-1b}
P^{(0)}_{n,l;\,n_3,l} = \frac{u^{n+1}}{(2l+1)!}\,
\frac{ (w-1)^{n_3-n-1}}{w^{n+n_3+3}}\, [n_3-n+ 2(n+1)(1-w)]\,z^{l-n} \,
{}_2F_1 (l-n, l-n_3; 2l+2; z).
\end{multline}
In the case of $l_2=1$ there are two possibilities.
Since the combination $l+l_2+l_3$ must be an even number, we have that
$l_3=l\pm 1$.
Below we calculate the coefficient $P^{(1)}_{n,l;\,n_3,l-1} (u,w)$.
(The coefficient $P^{(1)}_{n,l-1;\,n_3,l}$ can be derived using the
symmetry relation (\ref{eq:p-symm-1}).)
The calculation procedure is similar to that considered above.
As a result, we have,
\begin{multline}
  \label{eq:dip-3}
P^{(1)}_{n,l;\,n_3,l-1} = \frac{u^{n_3+2}}{6\,(2l+1)!}
\deriv_w w^{-n-n_3-3}  (w-u)^{n-n_3-1} \deriv^2_t t^{n_3+l+2}  \\
%%%%%%%%%%%%%%%%%%%%%%%
\times
 z^{l-1-n_3}\, {}_2F_1 (l-n, l-1-n_3; 2l+2; z) \Bigl|_{t=1},
\end{multline}
where $z=u t/[(w-u)(w-t)]$.
Omitting details of somewhat lengthy calculations we present only the final
result which is given by eq.~(\ref{eq:dip-5}) of
sec.~\ref{sec:matrix-elements}.

%%%%%%%%%%%%%%%%%%%%%%%%%%%%%%%

\section{Recurrence relations for the functions $P^{(l_2)}_{n,l;\,n_3,l_3}$}
\label{sec:recurrences}

The differential representation (\ref{eq:sum-3a}) can be used for the
derivation of the recurrence relations for the functions
$P^{(l_2)}_{n,l;\,n_3,l_3}$.
Such relations can be derived using the properties of the differential
operators.
For example, we note the following operator identity
\begin{multline}
  \label{eq:dif-1}
(n_3-l_3) \deriv^{n_3-1-l_3}_t
(1+t)^{n_3+l_3} + \deriv^{n_3-l_3}_t (1+wt)^{n_3+l_3}  \\
%%%%%%%%%%%%%%%
= (n_3+l_3+1) \deriv^{n_3-l_3-1}_t (1+wt)^{n_3+l_3} 
+ \deriv^{n_3-l_3-1}_t (1+wt)^{n_3+l_3+1} \deriv_t.
\end{multline}
(It can be proved by straightforward calculations.)
A similar identity can be written for the operators $\deriv_\tau$.
These two identities result into a pair of recurrences for the functions
$P^{(l_2)}_{n,l;\,n_3,l_3}$ with fixed multipole index $l_2$,
\begin{equation}
  \label{eq:rec-1}
  \begin{split}
    (n_3+l_3+1)\, P_{n,l;\,n_3,l_3} - (n_3-l_3)\, P_{n,l;\,n_3-1,l_3}
&= P_{n,l;\,n_3,l_3-1} - P_{n,l;\,n_3-1,l_3-1}, \\
%%%%%%%%%%%%%%%%%%%%%%
(n+l+1)\, P_{n,l;\,n_3,l_3} - (n-l)\, P_{n-1,l;\,n_3,l_3}
&= P_{n,l-1;\,n_3,l_3} - P_{n-1,l-1;\,n_3,l_3},
  \end{split}
\end{equation}
where we have omitted the superscript $l_2$ for the sake of shortness.
The recursion connecting functions with different indices $l_2$ can be derived
using the differential identity,
\begin{equation}
  \label{eq:r-6}
  \deriv^{l-l_2+l_3+2}_\tau R_{l_2}
=  2 (l_2+1) \bigl[(\xi-t-\tau) \deriv^{l-l_2+l_3+1}_\tau R_{l_2+1}
-  (l-l_2+l_3+1) \deriv^{l-l_2+l_3}_\tau R_{l_2+1} \bigr],
\end{equation}
where $R_{l_2} = [1- 2 \xi (t+\tau)+(t+\tau)^2]^{-l_2-1}$ and
$\xi=-(1+u)/(2w)$.
After some transformations, the above identity leads to the recursion 
connecting functions $P$ with indices $l_2$, $l_2+1$,
\begin{multline}
  \label{eq:l2-rec-2}
(n+l+2)\, P^{(l_2)}_{n+1, n_3} + (n-l)\, P^{(l_2)}_{n-1, n_3}
- 2 (n+1) P^{(l_2)}_{n, n_3}
=  (l_2+1) \Bigl(
 (n+l+2) (1-u)\, P^{(l_2+1)}_{n+1,n_3} \\
%%%%%%%%%%%%%%%
+ (n-l) (1+u) \, P^{(l_2+1)}_{n-1,n_3}
+ 2 [u(n_3+l_2+2)-n-1]\,P^{(l_2+1)}_{n,n_3}
- 2 (n_3+l_3+2) u\, P^{(l_2+1)}_{n,n_3+1} \Bigr).
\end{multline}
where $P^{l_2}_{n,n_3} = P^{(l_2)}_{n,l;\,n_3,l_3}$, \textit{etc}.
The recursion connecting $P$-functions with fixed values of angular momentum
indices $l,l_2,l_3$ can be derived using the fact that 
$\deriv_\tau R_{l_2} = \deriv_t R_{l_2}$.
Omitting detail of the routine transformations, we present only the final
result,
\begin{multline}
  \label{eq:rec-fix-n}
  (n_3+l_3+2)\, u P_{n,n_3+1} + (n_3-l_3)\,u P_{n, n_3-1} 
 - (n+l+2)\, P_{n+1, n_3} - (n-l)\, P_{n-1, n_3}\\
%%%%%%%%%%%%%%%%
+ 2 [n+1-u(n_3+1)]\, P_{n, n_3}  =0,
\end{multline}
where 
%% we have omitted the angular momentum indices in the notations,
$P_{n,n_3} \equiv P^{(l_2)}_{n,l;\, n_3, l_3}$.

Finally, we note that there are two other recurrence relations which connect
functions $P$ with different values of indices $n$, $n_3$.
Both of those recursions contain eight terms and they form a complete
set with respect to the indices $n$ and $n_3$.
It means that having 10 ``neighbor'' parameters one can recursively
calculate the parameters $P_{n,n_3}$ with arbitrary $n$ and $n_3$.
However, since these recursions have quite cumbersome form, we do not present
them here.

%%%%%%%%%%%%%%%%%%

%% \bibliographystyle{unsrt}
%% \bibliographystyle{alpha}
%% \bibliographystyle{apsrev}
%% \bibliography{mybib,Aquilanti,Avery_w_abstracts}

\begin{thebibliography}{38}
\expandafter\ifx\csname natexlab\endcsname\relax\def\natexlab#1{#1}\fi
\expandafter\ifx\csname bibnamefont\endcsname\relax
  \def\bibnamefont#1{#1}\fi
\expandafter\ifx\csname bibfnamefont\endcsname\relax
  \def\bibfnamefont#1{#1}\fi
\expandafter\ifx\csname citenamefont\endcsname\relax
  \def\citenamefont#1{#1}\fi
\expandafter\ifx\csname url\endcsname\relax
  \def\url#1{\texttt{#1}}\fi
\expandafter\ifx\csname urlprefix\endcsname\relax\def\urlprefix{URL }\fi
\providecommand{\bibinfo}[2]{#2}
\providecommand{\eprint}[2][]{\url{#2}}

\bibitem[{\citenamefont{Fock}(1935)}]{fock35:_o4}
\bibinfo{author}{\bibfnamefont{W.}~\bibnamefont{Fock}}, \bibinfo{journal}{Z.
  Phys.} \textbf{\bibinfo{volume}{98}}, \bibinfo{pages}{145}
  (\bibinfo{year}{1935}).

\bibitem[{\citenamefont{Lieber}(1968)}]{lieber68:_o4_lamb}
\bibinfo{author}{\bibfnamefont{M.}~\bibnamefont{Lieber}},
  \bibinfo{journal}{Phys. Rev.} \textbf{\bibinfo{volume}{174}},
  \bibinfo{pages}{2037} (\bibinfo{year}{1968}).

\bibitem[{\citenamefont{Bander and
  Itzykson}(1966)}]{bander_itzykson66I:_o4_hydrog}
\bibinfo{author}{\bibfnamefont{M.}~\bibnamefont{Bander}} \bibnamefont{and}
  \bibinfo{author}{\bibfnamefont{C.}~\bibnamefont{Itzykson}},
  \bibinfo{journal}{Rev. Mod. Phys.} \textbf{\bibinfo{volume}{38}},
  \bibinfo{pages}{330} (\bibinfo{year}{1966}).

\bibitem[{\citenamefont{Schwinger}(1964)}]{schwinger64:_c_green_f}
\bibinfo{author}{\bibfnamefont{J.}~\bibnamefont{Schwinger}},
  \bibinfo{journal}{J. Math. Phys.} \textbf{\bibinfo{volume}{5}},
  \bibinfo{pages}{1606} (\bibinfo{year}{1964}).

\bibitem[{\citenamefont{Maquet}(1977)}]{maquet77:_green_fock}
\bibinfo{author}{\bibfnamefont{A.}~\bibnamefont{Maquet}},
  \bibinfo{journal}{Phys. Rev. A} \textbf{\bibinfo{volume}{15}},
  \bibinfo{pages}{1088} (\bibinfo{year}{1977}).

\bibitem[{\citenamefont{Shibuya and
  Wulfman}(1965)}]{shibuya_wulfman65:_mo_fock}
\bibinfo{author}{\bibfnamefont{T.}~\bibnamefont{Shibuya}} \bibnamefont{and}
  \bibinfo{author}{\bibfnamefont{C.~E.} \bibnamefont{Wulfman}},
  \bibinfo{journal}{Proc Roy Soc A} \textbf{\bibinfo{volume}{286}},
  \bibinfo{pages}{376} (\bibinfo{year}{1965}).

\bibitem[{\citenamefont{Avery}(2004)}]{avery04:_shibuya_wulfman_fock}
\bibinfo{author}{\bibfnamefont{J.}~\bibnamefont{Avery}}, \bibinfo{journal}{Int.
  J. Quantum Chem.} \textbf{\bibinfo{volume}{100}}, \bibinfo{pages}{121}
  (\bibinfo{year}{2004}).

\bibitem[{\citenamefont{Avery}(2000)}]{avery00:_hspher_book}
\bibinfo{author}{\bibfnamefont{J.}~\bibnamefont{Avery}},
  \emph{\bibinfo{title}{Hyperspherical Harmonics and Generalized Sturmians}},
  vol.~\bibinfo{volume}{4} of \emph{\bibinfo{series}{Progress in Theoretical
  Chemistry and Physics}} (\bibinfo{publisher}{Kluwer Academic Publishers},
  \bibinfo{address}{Dordrecht}, \bibinfo{year}{2000}).

\bibitem[{\citenamefont{Aquilanti et~al.}(1998)\citenamefont{Aquilanti,
  Cavalli, and Coletti}}]{aquilanti98prl:_o4_recouplings}
\bibinfo{author}{\bibfnamefont{V.}~\bibnamefont{Aquilanti}},
  \bibinfo{author}{\bibfnamefont{S.}~\bibnamefont{Cavalli}}, \bibnamefont{and}
  \bibinfo{author}{\bibfnamefont{C.}~\bibnamefont{Coletti}},
  \bibinfo{journal}{Phys. Rev. Lett.} \textbf{\bibinfo{volume}{80}},
  \bibinfo{pages}{3209} (\bibinfo{year}{1998}).

\bibitem[{\citenamefont{Aquilanti et~al.}(2003)\citenamefont{Aquilanti,
  Caligiana, Cavalli, and Coletti}}]{AquilantiCCC03}
\bibinfo{author}{\bibfnamefont{V.}~\bibnamefont{Aquilanti}},
  \bibinfo{author}{\bibfnamefont{A.}~\bibnamefont{Caligiana}},
  \bibinfo{author}{\bibfnamefont{S.}~\bibnamefont{Cavalli}}, \bibnamefont{and}
  \bibinfo{author}{\bibfnamefont{C.}~\bibnamefont{Coletti}},
  \bibinfo{journal}{Int. J. Quantum Chem.} \textbf{\bibinfo{volume}{92}},
  \bibinfo{pages}{212} (\bibinfo{year}{2003}).

\bibitem[{\citenamefont{Aquilanti et~al.}(2001)\citenamefont{Aquilanti,
  Cavalli, Coletti, di~Domenico, and
  Grossi}}]{aquilanti01:_rev_hypsp_mom_space}
\bibinfo{author}{\bibfnamefont{V.}~\bibnamefont{Aquilanti}},
  \bibinfo{author}{\bibfnamefont{S.}~\bibnamefont{Cavalli}},
  \bibinfo{author}{\bibfnamefont{C.}~\bibnamefont{Coletti}},
  \bibinfo{author}{\bibfnamefont{D.}~\bibnamefont{di~Domenico}},
  \bibnamefont{and} \bibinfo{author}{\bibfnamefont{G.}~\bibnamefont{Grossi}},
  \bibinfo{journal}{Int. Reviews in Physical Chemistry}
  \textbf{\bibinfo{volume}{20}}, \bibinfo{pages}{673} (\bibinfo{year}{2001}).

\bibitem[{\citenamefont{Aquilanti et~al.}(1997)\citenamefont{Aquilanti,
  Cavalli, and Coletti}}]{AquilantiCC97}
\bibinfo{author}{\bibfnamefont{V.}~\bibnamefont{Aquilanti}},
  \bibinfo{author}{\bibfnamefont{S.}~\bibnamefont{Cavalli}}, \bibnamefont{and}
  \bibinfo{author}{\bibfnamefont{C.}~\bibnamefont{Coletti}},
  \bibinfo{journal}{Chem. Phys.} \textbf{\bibinfo{volume}{214}},
  \bibinfo{pages}{1} (\bibinfo{year}{1997}).

\bibitem[{\citenamefont{Aquilanti et~al.}(1996)\citenamefont{Aquilanti,
  Cavalli, Coletti, and Grossi}}]{AquilantiCCG96}
\bibinfo{author}{\bibfnamefont{V.}~\bibnamefont{Aquilanti}},
  \bibinfo{author}{\bibfnamefont{S.}~\bibnamefont{Cavalli}},
  \bibinfo{author}{\bibfnamefont{C.}~\bibnamefont{Coletti}}, \bibnamefont{and}
  \bibinfo{author}{\bibfnamefont{G.}~\bibnamefont{Grossi}},
  \bibinfo{journal}{Chem. Phys.} \textbf{\bibinfo{volume}{209}},
  \bibinfo{pages}{405} (\bibinfo{year}{1996}).

\bibitem[{\citenamefont{Avery and Avery}(2004)}]{AveryA04}
\bibinfo{author}{\bibfnamefont{J.}~\bibnamefont{Avery}} \bibnamefont{and}
  \bibinfo{author}{\bibfnamefont{J.}~\bibnamefont{Avery}}, \bibinfo{journal}{J.
  Phys. Chem. A} \textbf{\bibinfo{volume}{108}}, \bibinfo{pages}{8848}
  (\bibinfo{year}{2004}).

\bibitem[{\citenamefont{Avery and Shim}(2001)}]{AveryS01}
\bibinfo{author}{\bibfnamefont{J.}~\bibnamefont{Avery}} \bibnamefont{and}
  \bibinfo{author}{\bibfnamefont{R.}~\bibnamefont{Shim}},
  \bibinfo{journal}{Int. J. Quantum Chem.} \textbf{\bibinfo{volume}{83}},
  \bibinfo{pages}{1} (\bibinfo{year}{2001}).

\bibitem[{\citenamefont{Avery}(1997)}]{Avery97}
\bibinfo{author}{\bibfnamefont{J.}~\bibnamefont{Avery}}, \bibinfo{journal}{J.
  Math. Chem.} \textbf{\bibinfo{volume}{21}}, \bibinfo{pages}{285}
  (\bibinfo{year}{1997}).

\bibitem[{\citenamefont{Avery and Hansen}(1996)}]{AveryH96}
\bibinfo{author}{\bibfnamefont{J.}~\bibnamefont{Avery}} \bibnamefont{and}
  \bibinfo{author}{\bibfnamefont{T.~B.} \bibnamefont{Hansen}},
  \bibinfo{journal}{Int. J. Quantum Chem.} \textbf{\bibinfo{volume}{60}},
  \bibinfo{pages}{201} (\bibinfo{year}{1996}).

\bibitem[{\citenamefont{Avery et~al.}(1996)\citenamefont{Avery, Hansen, Wang,
  and Antonsen}}]{AveryHWA96}
\bibinfo{author}{\bibfnamefont{J.}~\bibnamefont{Avery}},
  \bibinfo{author}{\bibfnamefont{T.~B.} \bibnamefont{Hansen}},
  \bibinfo{author}{\bibfnamefont{M.~C.} \bibnamefont{Wang}}, \bibnamefont{and}
  \bibinfo{author}{\bibfnamefont{F.}~\bibnamefont{Antonsen}},
  \bibinfo{journal}{Int. J. Quantum Chem.} \textbf{\bibinfo{volume}{57}},
  \bibinfo{pages}{401} (\bibinfo{year}{1996}).

\bibitem[{\citenamefont{Muljarov et~al.}(2000)\citenamefont{Muljarov,
  Yablonskii, Tikhodeev, Bulatov, and Birman}}]{muljarov00jmp:_exciton_fock}
\bibinfo{author}{\bibfnamefont{E.~A.} \bibnamefont{Muljarov}},
  \bibinfo{author}{\bibfnamefont{A.~L.} \bibnamefont{Yablonskii}},
  \bibinfo{author}{\bibfnamefont{S.~G.} \bibnamefont{Tikhodeev}},
  \bibinfo{author}{\bibfnamefont{A.~E.} \bibnamefont{Bulatov}},
  \bibnamefont{and} \bibinfo{author}{\bibfnamefont{J.~L.}
  \bibnamefont{Birman}}, \bibinfo{journal}{J. Math. Phys.}
  \textbf{\bibinfo{volume}{41}}, \bibinfo{pages}{6026} (\bibinfo{year}{2000}).

\bibitem[{\citenamefont{Aquilanti and Caligiana}(2002)}]{AquilantiC02}
\bibinfo{author}{\bibfnamefont{V.}~\bibnamefont{Aquilanti}} \bibnamefont{and}
  \bibinfo{author}{\bibfnamefont{A.}~\bibnamefont{Caligiana}},
  \bibinfo{journal}{Chem. Phys. Lett.} \textbf{\bibinfo{volume}{366}},
  \bibinfo{pages}{157} (\bibinfo{year}{2002}).

\bibitem[{\citenamefont{Bethe}(1930)}]{bethe30:_GOS_passage_rapid_particles}
\bibinfo{author}{\bibfnamefont{H.}~\bibnamefont{Bethe}}, \bibinfo{journal}{Ann.
  Physik} \textbf{\bibinfo{volume}{5}}, \bibinfo{pages}{325}
  (\bibinfo{year}{1930}).

\bibitem[{\citenamefont{Inokuti}(1971)}]{inokuti71:_GOS_Bethe_inelastic}
\bibinfo{author}{\bibfnamefont{M.}~\bibnamefont{Inokuti}},
  \bibinfo{journal}{Rev. Mod. Phys.} \textbf{\bibinfo{volume}{43}},
  \bibinfo{pages}{297} (\bibinfo{year}{1971}).

\bibitem[{\citenamefont{Mott and Massey}(1965)}]{mott_massey65}
\bibinfo{author}{\bibfnamefont{N.~F.} \bibnamefont{Mott}} \bibnamefont{and}
  \bibinfo{author}{\bibfnamefont{H.~S.~W.} \bibnamefont{Massey}},
  \emph{\bibinfo{title}{The theory of atomic collisions}},
  vol.~\bibinfo{volume}{II} (\bibinfo{publisher}{Oxford University Press},
  \bibinfo{year}{1965}).

\bibitem[{\citenamefont{Jetzke and Broad}(1986)}]{jetzke86:_boost_hydrogen}
\bibinfo{author}{\bibfnamefont{S.}~\bibnamefont{Jetzke}} \bibnamefont{and}
  \bibinfo{author}{\bibfnamefont{J.~T.} \bibnamefont{Broad}},
  \bibinfo{journal}{J. Phys. B: At. Mol. Phys.} \textbf{\bibinfo{volume}{19}},
  \bibinfo{pages}{L199} (\bibinfo{year}{1986}).

\bibitem[{\citenamefont{Parzy\'nski and
  Sobczak}(2003)}]{parzynski03:_h_mel_nondip}
\bibinfo{author}{\bibfnamefont{R.}~\bibnamefont{Parzy\'nski}} \bibnamefont{and}
  \bibinfo{author}{\bibfnamefont{M.}~\bibnamefont{Sobczak}},
  \bibinfo{journal}{Opt. Comm.} \textbf{\bibinfo{volume}{225}}
  (\bibinfo{year}{2003}).

\bibitem[{\citenamefont{Holt}(1969)}]{holt69:_hydrogen_mel_bound_free}
\bibinfo{author}{\bibfnamefont{A.~R.} \bibnamefont{Holt}}, \bibinfo{journal}{J.
  Phys. B: At. Mol. Phys.} \textbf{\bibinfo{volume}{2}}, \bibinfo{pages}{1209}
  (\bibinfo{year}{1969}).

\bibitem[{\citenamefont{Datta}(1985)}]{datta85:_coulom_integrals}
\bibinfo{author}{\bibfnamefont{S.}~\bibnamefont{Datta}}, \bibinfo{journal}{J.
  Phys. B: At. Mol. Phys} \textbf{\bibinfo{volume}{18}}, \bibinfo{pages}{853}
  (\bibinfo{year}{1985}).

\bibitem[{\citenamefont{Meremianin}(2006)}]{avm05:_o4_jpa}
\bibinfo{author}{\bibfnamefont{A.~V.} \bibnamefont{Meremianin}},
  \bibinfo{journal}{J. Phys. A: Math. Gen.} \textbf{\bibinfo{volume}{39}},
  \bibinfo{pages}{3099} (\bibinfo{year}{2006}).

\bibitem[{\citenamefont{Manakov et~al.}(1996)\citenamefont{Manakov, Marmo, and
  Meremianin}}]{Man-96}
\bibinfo{author}{\bibfnamefont{N.~L.} \bibnamefont{Manakov}},
  \bibinfo{author}{\bibfnamefont{S.~I.} \bibnamefont{Marmo}}, \bibnamefont{and}
  \bibinfo{author}{\bibfnamefont{A.~V.} \bibnamefont{Meremianin}},
  \bibinfo{journal}{J. Phys. B: At. Mol. Opt. Phys.}
  \textbf{\bibinfo{volume}{29}}, \bibinfo{pages}{2711} (\bibinfo{year}{1996}).

\bibitem[{\citenamefont{Meremianin and Briggs}(2003)}]{meremianin03:_phys_rep}
\bibinfo{author}{\bibfnamefont{A.~V.} \bibnamefont{Meremianin}}
  \bibnamefont{and} \bibinfo{author}{\bibfnamefont{J.~S.}
  \bibnamefont{Briggs}}, \bibinfo{journal}{Phys. Rep.}
  \textbf{\bibinfo{volume}{384}}, \bibinfo{pages}{121} (\bibinfo{year}{2003}).

\bibitem[{\citenamefont{Varshalovich et~al.}(1988)\citenamefont{Varshalovich,
  Moskalev, and Khersonskii}}]{Varsh}
\bibinfo{author}{\bibfnamefont{D.~A.} \bibnamefont{Varshalovich}},
  \bibinfo{author}{\bibfnamefont{A.~N.} \bibnamefont{Moskalev}},
  \bibnamefont{and} \bibinfo{author}{\bibfnamefont{V.~K.}
  \bibnamefont{Khersonskii}}, \emph{\bibinfo{title}{Quantum theory of angular
  momentum}} (\bibinfo{publisher}{World Scientific},
  \bibinfo{address}{Singapore}, \bibinfo{year}{1988}).

\bibitem[{\citenamefont{Erdelyi
  et~al.}(1953{\natexlab{a}})\citenamefont{Erdelyi, Magnus, Oberhettinger, and
  Tricomi}}]{Bateman-I}
\bibinfo{author}{\bibfnamefont{A.}~\bibnamefont{Erdelyi}},
  \bibinfo{author}{\bibfnamefont{W.}~\bibnamefont{Magnus}},
  \bibinfo{author}{\bibfnamefont{F.}~\bibnamefont{Oberhettinger}},
  \bibnamefont{and} \bibinfo{author}{\bibfnamefont{F.~G.}
  \bibnamefont{Tricomi}}, \emph{\bibinfo{title}{Higher transcendental
  functions}}, vol.~\bibinfo{volume}{I} (\bibinfo{publisher}{McGraw-hill book
  company, Inc}, \bibinfo{year}{1953}{\natexlab{a}}).

\bibitem[{\citenamefont{Miller and
  Platzman}(1957)}]{miller_platzman57:_inelast_collisions_he_GOS}
\bibinfo{author}{\bibfnamefont{W.~F.} \bibnamefont{Miller}} \bibnamefont{and}
  \bibinfo{author}{\bibfnamefont{R.~L.} \bibnamefont{Platzman}},
  \bibinfo{journal}{Proc. Phys. Soc. A} \textbf{\bibinfo{volume}{70}},
  \bibinfo{pages}{299} (\bibinfo{year}{1957}).

\bibitem[{\citenamefont{Gordon}(1929)}]{gordon29}
\bibinfo{author}{\bibfnamefont{W.}~\bibnamefont{Gordon}},
  \bibinfo{journal}{Ann. Phys.} \textbf{\bibinfo{volume}{2}},
  \bibinfo{pages}{1031} (\bibinfo{year}{1929}).

\bibitem[{\citenamefont{Krylovetsky et~al.}(2001)\citenamefont{Krylovetsky,
  Manakov, and Marmo}}]{manakov_jetp01:_free_param_cfg}
\bibinfo{author}{\bibfnamefont{A.~A.} \bibnamefont{Krylovetsky}},
  \bibinfo{author}{\bibfnamefont{N.~L.} \bibnamefont{Manakov}},
  \bibnamefont{and} \bibinfo{author}{\bibfnamefont{S.~I.} \bibnamefont{Marmo}},
  \bibinfo{journal}{Journal of Experimental and Theoretical Physics}
  \textbf{\bibinfo{volume}{92}}, \bibinfo{pages}{37} (\bibinfo{year}{2001}).

\bibitem[{\citenamefont{Manakov et~al.}(2004)\citenamefont{Manakov, Marmo, and
  Pronin}}]{manakov_jetp04:_cfg_hyperpolarizability}
\bibinfo{author}{\bibfnamefont{N.~L.} \bibnamefont{Manakov}},
  \bibinfo{author}{\bibfnamefont{S.~I.} \bibnamefont{Marmo}}, \bibnamefont{and}
  \bibinfo{author}{\bibfnamefont{E.~A.} \bibnamefont{Pronin}},
  \bibinfo{journal}{Journal of Experimental and Theoretical Physics}
  \textbf{\bibinfo{volume}{98}}, \bibinfo{pages}{254} (\bibinfo{year}{2004}).

\bibitem[{\citenamefont{Hey}(2006)}]{hey06:h_matr_elem}
\bibinfo{author}{\bibfnamefont{J.~D.} \bibnamefont{Hey}}, \bibinfo{journal}{J.
  Phys. B: At. Mol. Opt. Phys.} \textbf{\bibinfo{volume}{39}},
  \bibinfo{pages}{2641} (\bibinfo{year}{2006}).

\bibitem[{\citenamefont{Erdelyi
  et~al.}(1953{\natexlab{b}})\citenamefont{Erdelyi, Magnus, Oberhettinger, and
  Tricomi}}]{Bateman-II}
\bibinfo{author}{\bibfnamefont{A.}~\bibnamefont{Erdelyi}},
  \bibinfo{author}{\bibfnamefont{W.}~\bibnamefont{Magnus}},
  \bibinfo{author}{\bibfnamefont{F.}~\bibnamefont{Oberhettinger}},
  \bibnamefont{and} \bibinfo{author}{\bibfnamefont{F.~G.}
  \bibnamefont{Tricomi}}, \emph{\bibinfo{title}{Higher trancendental functions.
  Bateman manuscript project}}, vol.~\bibinfo{volume}{II}
  (\bibinfo{publisher}{McGraw-hill book company, Inc},
  \bibinfo{year}{1953}{\natexlab{b}}).

\end{thebibliography}

\end{document}